# Survival of the metallic state in a single-hole multiband *p*-orbital molecular system

Keisuke Matsui[1#], Ryan A. Klein[2,3#], Naoya Yoshikane[4], John Arvanitidis[5], Matjaž Gomilšek[6], Urh Klopčič[6], Shogo Kawaguchi[7], Hitoshi Yamaoka[8], Nozomu Hiraoka[9], Hirofumi Ishii[9], Qiang Zhang[10], Shigeo Mori[1], Hiroki Ishibashi[4], Yoshiki Kubota[4], Craig M. Brown[3*], Denis Arčon[6,11*] & Kosmas Prassides[4,6*]

[1] Department of Materials Science, Graduate School of Engineering, Osaka Metropolitan University, Osaka 599-8531, Japan

[2] Department of Chemistry, University of California, Berkeley, California 94720, United States

[3] Center for Neutron Research, National Institute of Standards and Technology (NIST), Gaithersburg, Maryland 20899, USA

[4] Department of Physics, Graduate School of Science, Osaka Metropolitan University, Osaka 558-8585, Japan

[5] Physics Department, Aristotle University of Thessaloniki, Thessaloniki 54124, Greece

[6] Institute "Jožef Stefan", Jamova 39, SI-1000 Ljubljana, Slovenia

[7] Japan Synchrotron Radiation Research Institute (JASRI), 1-1-1 Kouto, Sayo-cho, Sayo-gun, Hyogo 679-5198, Japan

[8] RIKEN SPring-8 Center, Sayo, Hyogo 679-5148, Japan

[9] National Synchrotron Radiation Research Center, Hsinchu 30076, Taiwan

[10] Neutron Scattering Division, Oak Ridge National Laboratory, Oak Ridge, Tennessee 37831, USA

[11] Faculty of Mathematics and Physics, University of Ljubljana, Jadranska 19, SI-1000 Ljubljana, Slovenia

[#] These authors contributed equally to the manuscript.

* e-mail: craig.brown@nist.gov; denis.arcon@ijs.si; kosmas.prassides@ijs.si

***Abstract***: Strong correlations and ferromagnetic Hund's coupling lead to diverse electronic phenomena in transition-metal oxides that sensitively depend on the *d*-orbital electron filling. Fullerides, their *p*-electron counterparts, exhibit effective antiferromagnetic Hund's coupling in a different energy range. At half-filling ($n$=3, three electrons in triply degenerate orbitals), both *d*- and *p*-electron systems are Mott insulators due to strong correlations and Hund's coupling. Away from half-filling, in single-electron/hole ($n$=1,5) *d*-orbital systems, Hund's coupling opposes the correlations, reducing the Mott gap and allowing survival of metallicity. Here we report a single-hole multiorbital correlated *p*-electron system, orthorhombic-structured $Yb_2CsC_{60}$ comprising pentavalent $C_{60}^{5-}$ anions, which also exhibits a robust metallic state with no Mott transition, just like in the metastable single-electron cubic-structured $CsC_{60}$. We assert that particle–hole symmetry holds well in ($n$=1,5) fullerides and that their *p*-electron-derived states are analogous to those in *d*-orbital solids, providing impetus for further study of these correlated systems.

## Introduction

The Hubbard model is a paradigmatic model that has been extensively applied to materials whose properties are dominated by strong electron correlations.[1–3] Its two main parameters are the hopping integrals, $t$, principally between nearest-neighbour sites, and the on-site correlation energy, $U$. Despite its apparent simplicity, the model still captures intricate phenomena associated with complex $d$-orbital materials remarkably well, including the occurrence of Mott insulator-to-metal transitions, the emergence of unconventional superconductivity, and the appearance of magnetic phases.[2,3]

The application of the Hubbard model can be extended to multi-orbital systems, such as the Cr- or Ru-based oxides. In such correlated multiband 3$d$- and 4$d$-electron systems, the effect of Hund's rule coupling, $J_H$, on Mott localisation also needs to be considered.[4] For half-filled bands, *e.g.*, in the case of three electrons in three $t_{2g}$-orbitals, $J_H$ promotes electron correlation effects. On the other hand, when a single electron or hole is injected into the triply degenerate $t_{2g}$-orbitals, the effect of Hund's rule coupling is reversed: correlation effects are diminished, the Mott gap is reduced, and the critical value of the intra-orbital correlations, $U_c$, at which Mott localization takes place, increases. As a result, the metallic state is favoured and survives to larger values of $U$. For the case of two electrons or holes residing in the $t_{2g}$-orbitals, $J_H$ leads to a conflicting response in the electronic properties – the Mott gap is reduced while, at the same time, the degree of correlations increases. As a result, here $J_H$ promotes the emergence of "bad metallic behaviour".

This kind of interplay is much less understood both theoretically and experimentally for $p$-orbital molecular solids, which occupy different energy regimes from those of their $d$-orbital counterparts. An excellent $p$-orbital analogue of $d$-orbital systems is the family of fulleride molecular solids, which feature three electronically active $t_{1u}$ molecular orbitals.[5] In these materials, the bare Hund's coupling is weak,[6] and the effective sign of the exchange interaction, $J_{eff}$, is actually reversed into antiferromagnetic by the presence of Jahn-Teller phonons.[7] Theoretical studies of half-filled fullerides, $A_3C_{60}$ (A = alkali metal) have shown that the ground state is an antiferromagnetic Mott insulator,[7–9] as has indeed been experimentally observed for the most expanded member of this family, $Cs_3C_{60}$.[10–19] Application of physical and/or chemical pressure can tune the electron bandwidth, $W \sim 0.5$ eV, and thus the ratio $U/W$, leading to the stabilization of metallic and unconventional superconducting phases.[10,11,18,19] However, departing from half-filling of the $t_{1u}$ bands in a similar manner to that possible in multiband 3$d$ and 4$d$ electron systems has proven to be experimentally very challenging. The quenched cubic phase of $CsC_{60}$ (ref. [20]) is the only example of fullerides for which single-electron occupancy of the triply-degenerate $t_{1u}$-derived bands has been studied.[21,22] A temperature-independent electron paramagnetic resonance (EPR) spin susceptibility study of cubic $CsC_{60}$ at low temperatures suggests the adoption of a metallic ground state,[21] pointing to a possible parallel with the Hubbard model's prediction of robust metallic phases for single $d$-electron systems.[4] However, a detailed $^{133}$Cs nuclear magnetic resonance (NMR) study actually revealed the adoption of a more complex metallic state in which a small fraction of doubly-occupied $C_{60}^{2-}$ sites in their singlet state is encountered.[22] Such an unusual example of charge localization was attributed to the Jahn-Teller interaction, which opposes Coulomb repulsion and favours localized doubly-occupied $t_{1u}$ orbitals.

The behaviour of the Mott-insulating parent state of $A_3C_{60}$ and that of the metallic $AC_{60}$ phase seem to parallel that of *d*-orbital solids and agree with the predictions of the Hubbard model.[7–9] However, the critical case of single-hole $t_{1u}$-orbital fullerides for which particle-hole symmetry could be tested has yet to be addressed in detail, due to the lack of available model compounds. The search for new materials away from half-filling is even more urgent in light of recent theoretical predictions of superconductivity with a remarkably high critical temperature, $T_c$, at the quarter-filling level (and the three-quarter-filling level, if the particle-hole symmetry is not broken).[23] Previous attempts to tackle this part of the phase diagram concentrated on $Ba_2AC_{60}$,[24–26] for which a metallic state was suggested by the very small and weakly temperature-dependent EPR spin susceptibility, yet a final consensus on the ground state is still to be reached.

In this work, we address the electronic state of materials in which there is a single hole in the $t_{1u}$ orbitals by synthesizing and studying the $Yb_2CsC_{60}$ ternary fulleride. Our comprehensive structural and spectroscopic studies find that $Yb_2CsC_{60}$, in which the formal charge of the $C_{60}$ units is −5 – a single hole in the $t_{1u}$ orbitals – adopts a robust metallic state, *albeit* with a low density-of-states. The present experimental data, combined with first-principles calculations, enable the quantitative determination of the electronic bandwidth $W$, the non-zero crystal field resulting from crystal symmetry lowering, and the electronic density-of-states. These, together with the known value of the on-site correlation energy $U$, can rationalize why metallicity survives in the single-hole multiband *p*-orbital $Yb_2CsC_{60}$ system. Namely, Hund's exchange leads to the diminishing influence of correlations away from half-filling, indicating a parallel between *p*-electron molecular solids and analogous *d*-electron materials.

## Results and discussion

**Synthesis and crystal structure.** $Yb_2CsC_{60}$ powder samples were prepared by conventional solid-state reactions of appropriate pre-reacted fullerides (see Methods). Inspection of synchrotron X-ray (SXRPD) and neutron powder diffraction (NPD) patterns both at 100 K (Fig. S1) and room temperature (Fig. S2) readily reveals that $Yb_2CsC_{60}$ does not crystallize in the face-centred-cubic (*fcc*) $Fm\overline{3}m$ space group commonly adopted by ternary $A_2A'C_{60}$ fullerides (A, A' = alkali metals) nor in the $Fm\overline{3}$ space group of the $Ba_2CsC_{60}$ analogue (Fig. S3).[24,26] Instead, all reflections can be assigned either to a body-centred cell (space group *Immm*) with lattice parameters, $a_O \neq b_O \approx (1/\sqrt{2})a_C$ and $c_O \approx a_C$, where $a_C$ is the lattice parameter of the parent *fcc* cell, or to a small fraction of an $Yb_2O_3$ impurity. However, two additional very weak peaks, not attributable to an impurity phase, are clearly visible in the SXRPD profiles in the vicinity of $Q \sim 1.0$ and 1.36 Å$^{-1}$ (insets Fig. S2). These extra reflections, whose intensities are on the order of 3% of that of the weak (101) reflection, index as (111) and (021) on the orthorhombic unit cell. They provide an unambiguous signature of the violation of body-centred symmetry, i.e., the crystal symmetry reduction to a primitive orthorhombic cell (space group *Pmnn*). No other peaks that could arise from the lifting of the body-centred symmetry are observed under the present experimental conditions. The symmetry lowering to an orthorhombic unit cell is further supported by the observed strong splitting of the $H_g$ intramolecular $C_{60}$ modes in the Raman spectrum of $Yb_2CsC_{60}$ (Fig. S4). Temperature-dependent SXRPD and NPD measurements show that the orthorhombic unit cell smoothly contracts with decreasing temperature down to 10 K without the appearance of any new reflections or additional peak splitting, hence suggesting the absence of structural and/or magnetic phase transitions (Fig. S5).

**Crystal structure at 100 K.** Using the Pawley pattern-decomposition technique, a body-centred orthorhombic unit cell is thus taken as an initial step in our structural refinement. A structural refinement was undertaken by a combined Rietveld analysis of the SXRPD and NPD data at 100 K (Fig. S1). $Yb_2CsC_{60}$ crystallizes in the primitive orthorhombic *Pmnn* space group in which the ytterbium cations are allowed to shift along both the $y$ and $z$ directions. As a result, the two extra weak reflections are successfully reproduced with crystallographically distinct $Cs^+$ and $Yb^{2+}$ ions residing at the 2$c$ (0,0,½) and 4$g$ (½,$y$,$z$) Wyckoff sites (Fig. 1a, Table S1), respectively, in a unit cell of dimensions $a$ = 9.97392(15) Å, $b$ = 9.76416(16) Å, and $c$ = 13.8828(2) Å (values in parentheses indicate an experimental uncertainty of ±1 standard deviation $\sigma$). The oxidation state of the ytterbium ions (2+) and the resulting nominal charge of the fullerene anions (5−) are addressed in a later part of the manuscript.

Figure 1a illustrates schematically the derived structural model and its relationship to the parent *fcc* structure commonly found in the ternary $A_2A'C_{60}$ fullerides ($A$, $A'$ = alkali metals). Next, we consider the coordination environment of the metal ions. The larger $Cs^+$ cation forms contacts with the C–C fusions of six-membered rings (6:6 bonds, Fig. 1b) on two of the six adjacent $C_{60}$ units along the $c$-axis direction. Four additional contacts in the *ab* plane are with C–C fusions of hexagon-pentagon rings (6:5 bonds, Fig. 1b). The axial bonds as measured from the $Cs^+$ ion to the 6:6 bond centroids are $d$(Cs–6:6) = 3.433(7) Å, slightly shorter than the equatorial distances of $d$(Cs–6:5) = 3.494(5) Å in the *ab* plane, generating an axially compressed pseudo-octahedral environment. Conversely, the smaller $Yb^{2+}$ occupies a pseudo-tetrahedral interstitial site – it forms two unique contacts to C–C fusions of six- and five-membered rings on two adjacent $C_{60}$ units (6:5 bonds, $d$(Yb–6:5) = 2.703(5) Å and 2.787(5) Å) along the crystallographic <110> direction, and two identical contacts of 2.560(5) Å to the centroids of the pentagonal faces of the other two nearest-neighbouring $C_{60}$ anions along the <011> direction (Fig. 1c). The latter coordination environment of the $Yb^{2+}$ ion is reminiscent of that encountered in organometallic systems such as bis(allyl)bis(cyclopentadienyl)ytterbium,[27,28] which displays short Yb–C distances of *ca*. 2.6 Å to 2.7 Å,[29,30] smaller than the sum of the ionic radius of $Yb^{2+}$ (1.02 Å) and the van der Waals radius of C (1.70 Å) and indicative of covalent contributions to the bonding. Here, the Yb–C distances in $Yb_2CsC_{60}$ at 100 K are more extended, ranging from 2.821(5) to 2.865(5) Å, thereby implying the absence of any measurable covalency effects.

Given the orthorhombic crystal symmetry of $Yb_2CsC_{60}$ that leads to a reduction of the point group symmetry of the fullerene anions from $O_h$ in the *fcc* analogues to $C_{2h}$, we next investigate possible minor anisotropic distortions of the $C_{60}$ spherical shape. As a proxy, we measure the cross-$C_{60}$ distances between $C_{10}$, $C_{11}$, and $C_{15}$ atoms (Fig. S6, Table S4), yielding three cross-$C_{60}$ axes, $R_x$ ($x = a, b, c$) approximately along the $a$, $b$, and $c$ crystallographic directions, respectively. At 100 K, these cross-$C_{60}$ distances are 6.957(3) Å, 6.961(3) Å, and 7.045(18) Å, respectively, indicating the adoption of a prolate spheroidal molecular shape with the major axis along $c$ and with ($R_c/R_a$) and ($R_c/R_b$) ratios of 1.0127(5) and 1.0121(5), respectively. The distortion of the $C_{60}$ cage to $C_{2h}$ symmetry is driven by the static crystal field potential imposed on the molecular units by the anisotropic arrangement of the $Yb^{2+}$ and the $Cs^+$ ions. Even a relatively modest anisotropic electrostatic environment is sufficient to lift the icosahedral symmetry of the $C_{60}$ molecules, leading to small structural distortions, notably detectable in $Yb_2CsC_{60}$.

**Structural changes with temperature.** Temperature-dependent NPD patterns (Table S5, Fig. S5b, S8, S9) show that the orthorhombic unit cell anisotropically but smoothly contracts as the temperature decreases from room temperature down to 10 K (Fig. S7a), without any new reflections or significant changes in intensity that would suggest structural or magnetic phase transitions. Moreover, the unit cell volume per $C_{60}$, $V(T)$, monotonically decreases with temperature, but becomes temperature-independent at a value of 675.1(4) Å$^3$ below 25 K (Fig. S7b). The temperature dependence of $V(T)$ verifies the absence of any isostructural lattice anomaly due to electronic instabilities as encountered, *e.g.*, at the Mott insulator-to-metal transition in expanded $A_3C_{60}$ structures.[19] Interestingly, we observe a steeper slope for the change in $a$ with temperature compared to the change in $b$ or $c$, which appear similar to each other (Fig. 1d, Fig. S7a). To quantify this, we calculated the thermal expansion coefficients, $\alpha_x$ ($x = a, b, c$) between 50 K and 300 K from the NPD data as $\alpha_a = (\frac{1}{a})(\frac{da}{dT}) = 9.58(1)\times10^{-6}$ K$^{-1}$, which is significantly larger than $\alpha_b = 5.52(1)\times10^{-6}$ K$^{-1}$ and $\alpha_c = 5.79(1) \times 10^{-6}$ K$^{-1}$. The thermal expansion is thus anisotropic and positive with the largest value along the $a$ crystallographic direction.

Next, we sought to determine the structural origins of the anisotropic lattice expansion (Fig. 1d) by examining the subtle crystal structure changes deduced from the refinements of NPD data collected between 10 K and 300 K (Fig. S8, Fig. S9). We focus on the NPD patterns as a function of temperature because the neutron scattering lengths for naturally abundant C, Cs, and Yb are relatively similar at 6.64 fm, 5.42 fm, and 12.43 fm, respectively.[31] In contrast, the metal ions dominate the scattering in the SPXRD patterns, where the scattering is proportional to the number of electrons for each atom, such that the metal ions scatter the X-rays about an order of magnitude more than the C atoms. Overlaying the crystal structures obtained from Rietveld refinements against the NPD patterns collected at 10 K and 300 K,[32] there is almost no visible change in the position and orientation of the fulleride molecule, with bond lengths and angles consistent with idealized and literature values for fulleride systems across the temperatures (Fig. S10). Likewise, the $Cs^+$ ions occupy high-symmetry Wyckoff sites (2*c*) across the entire measured temperature range and so are structurally invariant with temperature.

Intriguingly, however, we observe a significant change in the Yb positions with temperature. Namely, in the temperature range of the present measurements, the $Yb^{2+}$ ions reside over the centre of the five-membered ring of the adjacent fullerene anions predominantly in the crystallographic *b*-axis direction (Fig. S10; colour of $Yb^{2+}$ ions is teal at 10 K and light blue at 300 K). To quantify this off-centring, we plot the *y* fractional coordinate of the Yb ion as a function of temperature (Fig. 1e) and contrast this with the distance from Yb to the centre of the pentagonal ring (Fig. 1f). We observe a monotonic increase in the Yb *y* coordinate as well as a monotonic decrease in the Yb–pentagon distance with decreasing temperature. There is an order of magnitude difference between the two shifts, with the Yb atom off-centring in the *b* direction being reduced by 0.117(9) Å between 300 and 10 K. In contrast, the Yb-pentagon distance decreases from 2.562(8) Å to 2.547(5) Å, a change of only 0.015(9) Å in the <101> direction. This is indicative of Yb staying essentially in the same position over the pentagonal ring. The Yb–(6:5) centroid distances elongate in the <011> (C13) direction (2.607(8) Å to 3.029(4) Å) and shrink towards C5 (2.900(8) Å to 2.799(4) Å) with decreasing temperature. The reduced off-centring shifts of the position of the $Yb^{2+}$ cations with decreasing temperature show that at the lowest temperatures, the $Yb_2CsC_{60}$ structure approaches, but does not fully achieve, a higher-symmetry body-centred *Immm* structure.

Simultaneously, we observe a decrease in the isotropic nuclear displacement parameters, $b_{eq}$ for Yb, Cs, and C on cooling. Fig. 1g plots the ratio of the isotropic displacement parameter, $b_{\mathrm{eq}}$, divided by the values obtained at 10 K to illustrate relative changes with temperature for each ion. Inspection of the relative changes for the $b_{\mathrm{eq}}$ values suggests significant dynamical motions that turn on at different temperatures for the Cs, Yb, and C atoms, following a rough trend in effective mass of the metal and fulleride ions (effective mass of the fulleride ion estimated as 720.66 amu, whereas the atomic masses of Cs and Yb are 132.91 amu 173.04 amu, respectively). The largest relative changes in isotropic displacement parameter are seen for Cs, in which $b_{eq}$ is initially invariant between 300 K and 250 K and then monotonically decreases below 250 K down to 10 K. The displacement parameter for the Yb atom is again invariant between 300 K and 250 K, followed by a monotonic decrease down to 25 K and a levelling off to 10 K. Interestingly, the C atom displacement parameter decreases sharply in the temperature interval between 300 K and 150 K but remains relatively invariant and small down to 10 K. The various changes in the relative $b_{eq}$ values are a proxy for dynamical ionic motion in the system and display variable dynamic behaviour for the constituent atoms in $Yb_2CsC_{60}$, with the lightest ion, Cs, being the most dynamic across the measured temperature interval. However, we note that, while the Cs ion displays the largest change relative to the 10 K value in the system, the $b_{eq}$ values themselves remain relatively small and in line with minimal rattling at 10 K ($b_{eq}$ = 0.20(14) Å$^2$) and moderate rattling at 300 K ($b_{eq}$ = 1.6(4) Å$^2$). Lastly, the cross-$C_{60}$ axes, $R_x$, are temperature invariant, implying that the spheroidal distortion of the $C_{60}$ units is robust over the entire temperature range. We can conclude that the fullerene anions likely undertake librational motions,[33] whereas the dynamical motions for the metal ions can be described as rattling-like behaviour.

***Electronic properties.*** The relatively close contact between Yb and the $C_{60}$ ions and the structural symmetry lowering to $C_{2h}$ of the latter may affect the nature of the electronic states, so we next address the oxidation state of the Yb atom. The $L_3$ X-ray absorption spectrum for $Yb_2CsC_{60}$ in the partial fluorescence yield (PFY-XAS) mode at room temperature shows a prominent peak at a photon energy of ~8939 eV, unambiguously signalling the existence of the Yb in its 2+ oxidation state (Fig. 2a).[34] Additional weak spectral intensity (~6% of the total signal) observable in the energy range between ~8943 and 8950 eV is characteristic of trivalent $Yb^{3+}$ cations associated with the presence of the $Yb_2O_3$ impurity phase, which has also been detected in the SXRPD and NPD datasets (Fig. S1, Table S1). The adoption of the oxidation state of 2+ for Yb is further supported by $^{171}Yb$ ($I$ = ½) NMR spectra measured between 290 K and 7 K (Fig. 2b), which comprise a single resonance peak attributed to the $Yb^{2+}$ cations in the pseudo-tetrahedral interstitial sites. No other signal in the range of frequencies anticipated for $Yb^{3+}$ (Ref. [35]) is detected within the resolution of this measurement.

With Yb adopting a 2+ oxidation state, the valence state of the $C_{60}$ anions in $Yb_2CsC_{60}$ must be 5−. This is consistent with the observed shift of the $^{13}C$ resonance in the room temperature magic angle spinning (MAS) NMR spectrum, which is offset from the tetramethylsilane (TMS) standard by 168 ppm (Fig. 2c). The measured $^{13}C$ NMR shift is far from those expected for $sp^3$ carbons, thus excluding any Yb–C covalent bonding and corroborating the conclusions from the structural analysis. In addition, the measured shift is larger than the chemical shifts of 143 ppm for $C_{60}$ (empty $t_{1u}$ orbitals)[36] and 156 ppm for hexavalent $C_{60}^{6-}$(fully occupied $t_{1u}$ orbitals).[37] The additional shift of $^{13}K_{hf}$ ~14 ppm from the chemical shifts of 154 ppm for $C_{60}^{5-}$ is attributed to the hyperfine coupling of $^{13}C$ nuclear spins to unpaired electrons in the $t_{1u}$ orbitals, and is similar in size to those measured in other metallic fulleride phases.[38,39] For comparison, in $Ba_2CsC_{60}$, two $^{13}C$ MAS NMR lines with shifts of 157.6 ppm and 177.3 ppm and with a 1:4 intensity ratio,[24,25] were observed, while a single $^{13}C$ NMR line shifted by ~155 ppm has been reported for the metallic phase of quenched cubic $CsC_{60}$ (note that the chemical shift for $C_{60}^{-}$ is 145 ppm).[40] The $C_{60}^{5-}$ valence state is further corroborated by Raman spectroscopic data for $Yb_2CsC_{60}$ at room temperature. Here we observe that the totally symmetric pentagonal-pinch $A_g(2)$ mode of $C_{60}$ is red-shifted by 26 $cm^{-1}$ relative to its position in pristine $C_{60}$ (Fig. 2d). Given that the frequency shift per one electron transferred to $C_{60}$ is ~5–6 $cm^{-1}$,[41,42] the observed frequency of the $A_g(2)$ mode in $Yb_2CsC_{60}$ lies precisely in the range expected for a $C_{60}^{5-}$ anion.

$^{13}C$ NMR provides a direct insight into the static and dynamic local spin susceptibility of the $C_{60}$ ions. At temperatures above ~310 K, the $^{13}C$ NMR spectra of $Yb_2CsC_{60}$ powder are very narrow. The fitting of the $^{13}C$ NMR spectrum measured at 340 K to a Lorentzian lineshape (Fig. S11) yields a $^{13}C$ NMR linewidth ($\delta\nu_{½}$) of 3.6 kHz. However, upon cooling, the $^{13}C$ NMR spectra start to broaden already at ~310 K (Fig. S12) and gradually develop a typical powder NMR lineshape. This line broadening is a clear hallmark of the slowing down of $C_{60}^{5-}$ molecular dynamics, with $C_{60}$ cages appearing static on the $^{13}C$ NMR timescale at low temperatures.[43] However, a careful examination of the low-temperature $^{13}C$ NMR spectra shows that the $^{13}C$ NMR lineshape simulation requires the introduction of another very weak component shifted by ~400 ppm relative to TMS to account for some missing signal intensity on the high-frequency side of the spectrum (inset to Fig. S12). The presence of two components in the $^{13}C$ NMR spectra is reminiscent of the situation in $Ba_2CsC_{60}$, where it was tentatively attributed to putative Jahn-Teller $C_{60}^{5-}$ distortions and to stronger coupling of C sites to the coordinating $Ba^{2+}$ ions in the tetrahedral sites. The distortion of $C_{60}^{5-}$ units to $C_{2h}$ symmetry in $Yb_2CsC_{60}$ as established by our structural analysis here provides support for a similar conclusion.

The centre of the $^{13}C$ NMR spectra (Fig. 2e) is very weakly temperature dependent. In general, the total shift of the $^{13}C$ NMR resonance has two contributions: the temperature-independent chemical shift and the $^{13}C$ hyperfine coupling ($^{13}K_{hf}$) to unpaired $t_{1u}$ electrons.[38] The latter is directly related to the spin-only susceptibility of $C_{60}^{5-}$ anions, $\chi(T)$ *via* the expression: $^{13}K_{\mathrm{hf}} = (^{13}a/N_{\mathrm{A}}\mu_{\mathrm{B}})\chi(T)$.[38,44] Here $^{13}a$ is the isotropic Fermi-contact hyperfine coupling constant, $\mu_{\mathrm{B}}$ the Bohr magneton, and $N_A$ the Avogadro number. A weak temperature dependence of the $^{13}C$ NMR shift thus signals a nearly-temperature-independent $\chi(T)$, providing evidence for a metallic state, similar to those encountered in other fulleride phases.[19,22,24] The weak temperature dependence of the $^{171}Yb$ and $^{133}Cs$ NMR shifts (Fig. 2e) supports the metallic state of $Yb_2CsC_{60}$ as well – the observed temperature dependence of both shifts is reminiscent of weakly temperature-dependent shifts of tetrahedral- and octahedral-site alkali-metals commonly observed in various metallic $A_3C_{60}$ salts.[19,38]

In metals, the $^{13}C$ spin-lattice relaxation rate, $1/^{13}T_1$, is expected to follow the Korringa relation, $1/^{13}T_1T = \frac{4\pi k_{\mathrm{B}}\gamma_{13}^2}{\hbar\gamma_{\mathrm{e}}^2}\frac{^{13}K_{hf}{}^2}{\beta} \propto g^2(E_F)$. Here, $\gamma_{13}$ and $\gamma_e$ are the $^{13}C$ nuclear and electronic gyromagnetic ratios, respectively, the phenomenological parameter $\beta$ is the so-called Korringa enhancement factor that characterizes the strength of electron correlations, and $g(E_F)$ is the electron density-of-states (DOS) at the Fermi level $E_F$.[45] In Fig. 2f, we show that $1/^{13}T_1T$ indeed becomes temperature independent below ~200 K and settles at a value of 0.93(6)×$10^{-3}$ $K^{-1}s^{-1}$, supporting the adoption of a metallic ground state in $Yb_2CsC_{60}$. The low-temperature $1/^{13}T_1T$ value is very close to those reported for the quenched cubic metallic phase of $CsC_{60}$ (~$10^{-3}$ $K^{-1}s^{-1}$)[40] and for $Ba_2CsC_{60}$ (~4×$10^{-3}$ $K^{-1}s^{-1}$).[24] Moreover, comparing the low-temperature $1/^{13}T_1T$ of $Yb_2CsC_{60}$ to typical values for $A_3C_{60}$ fullerides ($1/^{13}T_1T$ = 12×$10^{-3}$ $K^{-1}s^{-1}$),[19] we conclude that the density-of-states at the Fermi level, $g(E_F)$ in $Yb_2CsC_{60}$ is smaller by a factor of ~3 compared to $A_3C_{60}$ fullerides, assuming a similar Korringa enhancement factor, including $Na_2CsC_{60}$ which has a similar unit cell volume and a density-of-states at the Fermi level of ~15 states/eV/$C_{60}$ extracted from spin susceptibility data.[46,47] The $^{171}$Yb and $^{133}$Cs spin-lattice relaxation rates divided by temperature similarly show a very weak temperature dependence (Fig. 2f), providing further strong support for a metallic ground state.

The enhancement of $1/^{13}T_1T$ above ~200 K may be due to high-temperature $C_{60}^{5-}$ molecular dynamics. We find it intriguing that $1/^{13}T_1T$ vs $T$ (Fig. 2f) closely resembles the temperature dependence of the relative carbon isotropic displacement parameter, $b_{eq}$ (Fig. 1g), thus hinting a close link between the structural and the molecular charge/spin dynamics at higher temperatures (see supplementary information for a discussion). Alternatively, the higher temperature enhancement of $1/^{13}T_1T$ may indicate the presence of short-lived localized charge excitations into $C_{60}^{4-}$ and $C_{60}^{6-}$, in analogy to the charge disproportionation of $C_{60}^{3-}$ into $C_{60}^{2-}$ and $C_{60}^{4-}$ as seen in $Na_2CsC_{60}$.[48]

**Discussion.** Research on strongly correlated electronic states in $C_{60}$-derived solids has primarily focused on magnetic and superconducting $A_3C_{60}$ compositions, where the $t_{1u}$-derived bands are exactly half-filled (doping level, $n$ = 3). Away from half-filling, the metastable quenched cubic $CsC_{60}$ phase is an example of a metallic material with a single electron in the triply degenerate $t_{1u}$ orbitals ($n$ = 1), which turned out to show unusual electronic features due to the stabilisation of localised $C_{60}^{2-}$ singlets. However, accessing the properties of single-hole analogues (doping level, $n$ = 5 or equivalently, $h$ = 1) of the $n$ = 1 single-electron fullerides has proven very challenging, with all research reported to date restricted to the $Ba_2AC_{60}$ composition.[24–26] Here, we have successfully isolated the $Yb_2CsC_{60}$ composition as a potential isoelectronic and isostructural analogue. Given the frequent occurrence of 2+/3+ mixed valence in Yb compounds, including those of $C_{60}$,[49,50] we established unambiguously – using PFY-XAS, $^{171}$Yb NMR and Raman spectroscopic measurements (Fig. 2a–d) – that here the Yb oxidation state is +2, thereby unveiling $Yb_2CsC_{60}$ as an $h$ = 1 metal fulleride. On the other hand, combined SXRPD and NPD measurements reveal that $Yb_2CsC_{60}$ is not isostructural with $Ba_2CsC_{60}$. Substitution of the $Ba^{2+}$ ions ($r$ ~ 1.33 Å) residing in the tetrahedral interstices of the crystal structure by the smaller $Yb^{2+}$ ($r$ ~ 1.02 Å) ions leads to a significant reduction in the packing factor (~715 Å$^3$/$C_{60}$ in $Ba_2CsC_{60}$ *vs* ~680 Å$^3$/$C_{60}$ in $Yb_2CsC_{60}$). This is accompanied by a crystal symmetry reduction to primitive orthorhombic (space group *Pmnn*), where the $Yb^{2+}$ cations shift away from the centre of the tetrahedral interstices and the $C_{60}^{5-}$ ions reside in crystallographic sites of $C_{2h}$ symmetry (Fig. 1a, Fig.

S10). These crucial structural differences place $Yb_2CsC_{60}$ in a position on a fulleride phase diagram away from $Ba_2CsC_{60}$. Orthorhombic-structured $Yb_2CsC_{60}$ adopts a robust metallic state with a low density-of-states at the Fermi level, $g(E_F) \sim 5$ states/eV/$C_{60}$ extracted from the Korringa relation of the temperature dependence of the $^{13}$C NMR spin-lattice relaxation rate (Fig. 2f).

The $C_{60}^{n-}$ ($n$ = 1–5) anions are Jahn-Teller active molecular units in which coupling to $H_g$ phonon modes leads to distortions of the original high-symmetry $I_h$ molecular shape into structures with point groups $D_{2h}$, $D_{5h}$, or $D_{3d}$.[51,52] In orthorhombic $Yb_2CsC_{60}$, the $C_{60}^{5-}$ units adopt a prolate spheroidal shape (Fig. S6, Table S4), which is invariant over the temperature range of the present work, consistent with the existence of a cooperative static Jahn-Teller distortion of $C_{2h}$ symmetry. We note that the symmetry of the fulleride ion is lower than that of the Jahn-Teller-distorted molecular shape, implying that the final symmetry reduction to $C_{2h}$ is driven by crystal field effects.

In contrast to $Ba_2AC_{60}$, $Yb_2CsC_{60}$ exhibits no high-temperature structural and/or electronic instabilities. It crystallizes in a robust primitive orthorhombic unit cell and remains metallic throughout the temperature range of the present experiments. We stress that the $Ba_2AC_{60}$ crystal structures host orientational (merohedral) disorder of the $C_{60}$ anions. Conversely, $Yb_2CsC_{60}$ adopts a lower symmetry orthorhombic structure, but with orientationally ordered $C_{60}$ anions, thereby allowing us to uniquely characterize the subtle molecular distortions elusive in the $Ba_2AC_{60}$ analogues.

While both $Ba_2CsC_{60}$ and $Yb_2CsC_{60}$ are isoelectronic, the discovery of $Yb_2CsC_{60}$ enables investigations that were previously not possible. The synthetic breakthrough suggests that other members of the extended $M_2AC_{60}$ (M = Ba, Sr, Ca, Sm, Eu, Yb; $A$ = Cs, Rb, K, Na) family of materials should also become experimentally accessible. This would enable a systematic exploration of the structural and electronic phase diagram of pentavalent fullerides, analogous to what has been achieved for the trivalent $A_3C_{60}$ systems. The established robustness of metallicity in $n$ = 5 fullerides provides a unique opportunity to investigate the interplay between orbital degeneracy, electronic correlations away from half filling, and the evolution of Mott criticality with band filling. In particular, the absence of any instabilities renders $Yb_2CsC_{60}$ a particularly attractive platform for finely tuning electronic properties by controlling external stimuli such as pressure and substitutional doping, potentially enabling the emergence of superconductivity in strongly correlated non-cubic $t_{1u}$-based fullerides.

The structural and electronic findings are next addressed by *ab initio* density functional theory (DFT) calculations (see Methods). We first initiate the DFT calculations from the experimental *Pmnn* structure, but the structure spontaneously relaxes to the higher symmetry *Immm* space group. However, the energy difference between the two competing structures is minuscule – the computed difference in energy between the two structures is 138.7 meV/unit cell, equivalent to only ~1.1 meV/atom. We note that upon cooling to 10 K, the experimentally determined *Pmnn* structure tends towards the higher symmetry *Immm* structure (Fig. 1e), which is preferred by the DFT calculations implicitly performed at 0 K.

In the DFT band structure calculations (Fig. 3a), the energies of the $C_{60}$ $t_{1u}$-LUMO (lowest unoccupied molecular orbital) derived bands are clearly separated from those of the $C_{60}$ $h_u$-HOMO (highest occupied molecular orbital) derived bands and the $C_{60}$ $t_{1g}$–(LUMO+1) derived bands, although the band gaps appear smaller compared to the cubic $K_3C_{60}$ or the orthorhombic $(CH_3NH_2)K_3C_{60}$.[53] Not surprisingly, the $t_{1u}$-derived bands are the only ones that cross the Fermi energy, $E_F$, predicting a metallic state in agreement with experiment. The contribution from the Yb *d*- and *f*-orbitals to states at $E_F$ is negligibly small. The calculated density-of-states at $E_F$ is $g(E_F)$ = 5.4 states/eV/$C_{60}$ (Fig. 3a – right panel), and is thus more than three times smaller than $g(E_F)$ = 17 states/eV/$C_{60}$ calculated, for example, for the orthorhombic $(CH_3NH_2)K_3C_{60}$.[54] This *ab initio* prediction is also in good quantitative agreement with the experimental value of ~5 states/eV/$C_{60}$ reported here for $Yb_2CsC_{60}$ (Fig. 2f).

Because there are two $C_{60}$ molecules per unit cell in $Yb_2CsC_{60}$, six $t_{1u}$ orbitals contribute to the conduction band resulting in a total bandwidth of $W \sim 1$ eV (Fig. 3a). Taking a typical value for the on-site Coulomb interaction in fullerides of $U = 1$ eV,[8] we estimate $(U/W) \sim 1$; this is lower than the critical value $(U/W)_c$ for the onset of a metal-to-Mott insulator transition in fullerides (~2.3).[7–9,54] Single-hole ($h$ = 1) fulleride compounds thus indeed succeed in avoiding a metal-to-Mott insulator transition, which is reminiscent of the diminishing correlation effects of *d*-electrons in transition-metal oxides away from half filling.[4]

In the present band-structure calculations, we find a splitting of bands at the Γ point (Fig. 3a). The characteristic energy gain due to Jahn-Teller phonons is about 50 meV,[19,51] much less than the observed splitting of the bands in agreement with our earlier discussion that the crystal field is the leading cause of the stabilisation of the molecular distortion of the fullerene anions. At the same time, the crystal-field splitting of the three quasi-degenerate $t_{1u}$-derived bands is small compared to the electronic bandwidth and therefore unable to stabilize one or more of the three quasi-degenerate $t_{1u}$ orbitals. Inspecting more closely the bands that cross $E_F$, we find that four out of the six $t_{1u}$-derived bands are pushed towards slightly lower energies, leaving the other two to contribute their main DOS weight at $E_F$. A similar crystal field splitting of $C_{60}$ states was also discussed previously in the context of $Ba_2AC_{60}$ compounds[24] and was used to explain the observed splitting of the $^{13}C$ NMR peaks. The absence of such a clear $^{13}C$ NMR splitting here and the observation of only a low-intensity shoulder at high frequency (Fig. S12 inset), thus implies weaker crystal-field effects for $Yb_2CsC_{60}$. In fact, stronger crystal field effects would have eventually led to the isolation of a single half-filled $t_{1u}$-derived band with possible antiferromagnetic Mott instabilities due to correlation effects (Fig. 3b – top panel).[7] $Ba_2AC_{60}$ appears to be closer to this limit compared to $Yb_2CsC_{60}$, which may explain some of the fine differences between these two fulleride compositions, *e.g.*, the difference in the values of $1/^{13}T_1T$. From this perspective, $Yb_2CsC_{60}$ is actually electronically closer to the metastable cubic phase of $CsC_{60}$ with a single electron injected into $t_{1u}$-derived bands. Since both $Yb_2CsC_{60}$ and $CsC_{60}$ have nearly identical $1/^{13}T_1T$, we conclude that particle–hole symmetry seems to hold well in fullerides with $n = 1$ and $n = 5$ states. These π-electron-derived states encountered in fullerides are therefore directly analogous to those encountered in *d*-orbital solids, stimulating further search for correlated electronic behaviour in molecular solids. A remaining question concerns the importance of the effective Hund's exchange between the $t_{1u}$-derived states of $C_{60}^{5-}$, $J_{eff}$. Namely, strong Hund's coupling can give rise to a Hund's metal, a state which has been shown to bridge a high-spin Mott insulator and a charge-disproportionated insulator.[55] While a quantitative determination of $J_{eff}$ is beyond the scope of the present work, we note that fullerides reside in the very unusual regime of small, but negative, values of $J_{eff}$.[8,9,23] Even if $J_{eff}$ were to be enhanced by dynamical Jahn-Teller phonons, $Yb_2CsC_{60}$ in which $(U/W) \sim 1$ and $(J_{eff}/W)$ is small, should thus lie deep in the metallic region of the general fulleride phase diagram.[55] However, the band structure of $Yb_2CsC_{60}$ is consistent with (in total) two holes per two $C_{60}^{5-}$ injected into fully occupied $t_{1u}$-derived bands. Due to the overlap between the $t_{1u}$-derived sub-bands, the influence of effective Hund's exchange, $J_{eff}$, between these states may again become relevant, directly impacting the magnetic moments of $C_{60}^{5-}$ (Fig. 3b – bottom panel). This is an entirely unexplored situation in fullerides that has not yet been adequately theoretically addressed. Therefore, more experimental and theoretical work is needed to elucidate the role of $J_{eff}$ in the case of single-hole $t_{1u}$-derived bands and to investigate this part of the strongly correlated *p*-electron phase diagram.

**Methods**

**Sample preparation**

500-600 mg of $C_{60}$ (MER, 99.9%, sublimed) were first freshly sublimed in a dynamic vacuum (<5.0×$10^{-4}$ mbar) at 550 °C for about 16 h. After sublimation, $C_{60}$ was kept in an argon-filled glovebox ($O_2$ < 0.1 ppm, $H_2O$ < 0.2 ppm) or under other anaerobic conditions (helium-gas atmosphere or under high vacuum) for all reaction steps. Note: Certain commercial equipment, instruments, or materials are identified in this document. Such identification does not imply recommendation or endorsement by the National Institute of Standards and Technology, nor does it imply that the products identified are necessarily the best available for the purpose.

In the first step towards the synthesis of the targeted $Yb_2CsC_{60}$ ternary fulleride, two precursor compositions with stoichiometry $M_6C_{60}$ ($M$ = Cs, Yb) were prepared. Saturated caesium fulleride, $Cs_6C_{60}$ was synthesized by the reaction of sublimed $C_{60}$ with Cs vapour inside a Pyrex glass tube following procedures detailed elsewhere.[10–12,19] Laboratory powder X-ray diffraction patterns confirmed the high purity of the body-centred-structured (*bcc*)-$Cs_6C_{60}$ sample. In a similar way, the saturated Yb fulleride phase with nominal stoichiometry $Yb_6C_{60}$ was prepared by direct reaction of stoichiometric amounts of $C_{60}$ and Yb powder (Sigma-Aldrich, 99.9%). The $Yb_6C_{60}$ tablet was annealed at 400℃ for 3 h followed by a second heat treatment at 600℃ for 16 h to minimize the formation of Yb carbides at high temperatures. The powder XRD pattern of as-prepared $Yb_6C_{60}$ shows the presence of distinct *bcc*-indexable Bragg peaks. The as-prepared $Yb_6C_{60}$ sample was used as an Yb source for the synthesis of the targeted $Yb_2CsC_{60}$ composition in the next step. A pellet of the mixture of three precursors ("$Yb_6C_{60}$", $Cs_6C_{60}$, and pristine $C_{60}$ powder − 100 mg in total mass) was pressed and heated at ~400 °C for 4 h, followed by additional heating at ~600 °C for ~30 h. A second sample was prepared using Yb powder (Sigma-Aldrich, 99.9%), $Cs_6C_{60}$, and pristine $C_{60}$ and was thermally annealed using the same protocol. A sample for neutron diffraction experiments of ~450 mg in mass was prepared using the same thermal protocol but with a longer heat treatment at ~600℃ (3.5 days) to ensure the completion of the reaction and the homogenization of the larger amount of powder. Importantly, all samples consistently show the same powder XRD pattern, although they slightly vary in the amount of the $Yb_2O_3$ impurity.

As a reference material, we also synthesized a $Ba_2CsC_{60}$ sample. For this sample, the $Ba_6C_{60}$ precursor was prepared by solid-state reaction of stoichiometric amounts of Ba powder and $C_{60}$ at ~720 °C. Great care was taken to avoid contamination by oxidation by cleaning the surface of the purchased barium ingot (Sigma-Aldrich, >99.5%) using a diamond file, followed by grinding with another cleaned file. Still, the diffraction pattern of $Ba_6C_{60}$ shows the presence of a *bcc*-structured majority phase together with three minor impurities (BaO, $BaC_2$, and $Ba_3C_{60}$). In the next step, a mixture of stoichiometric amounts of three reactants ($Ba_6C_{60}$, $Cs_6C_{60}$, and pristine $C_{60}$ powder – 100 mg in total mass) was pressed into a pellet by using a set of stainless-steel dies (7 mm in diameter), which was then placed inside a tantalum tube. The tantalum tube was then sealed in a Pyrex glass (or quartz, for high-temperature annealing above 560 °C) tube (Ø15 mm) under 200 - 400 mbar of helium gas to improve thermal homogeneity during heat treatment. The sealed glass tube was then heated at ~400 °C for 4 h, followed by further heating at ~600 °C for ~30 h with two intermediate grindings. SXRPD confirmed the single-phase nature of the face-centred cubic (*fcc*)-$Ba_2CsC_{60}$ composition.

**Synchrotron X-ray powder diffraction (SXRPD)**

Polycrystalline $Yb_2CsC_{60}$ and $Ba_2CsC_{60}$ samples, each sealed in a standard glass capillary (0.3 mm outer diameter), were measured by SXRPD with the high-resolution powder diffractometer BL02B2 beamline, SPring-8, Japan ($\lambda$ = 0.79911(1) Å, calibrated with a $CeO_2$ standard, $a$ = 5.4111 Å) and equipped with a helium blower. Data for $Yb_2CsC_{60}$ were recorded at 30 K, 40 K, 50 K, 60 K, 70 K, 80 K, 90 K, 100 K, and 300 K. Data for $Ba_2CsC_{60}$ were recorded at 300 K.

**Neutron powder diffraction (NPD)**

Neutron diffraction measurements were carried out with time-of-flight (TOF) neutron powder diffractometer, POWGEN[56] (BL-11A) at the spallation neutron source (SNS) in the Oak Ridge National Laboratory, USA. Approximately 408 mg of polycrystalline $Yb_2CsC_{60}$ sample was loaded into a vanadium can with a 6 mm diameter inside a helium-filled glovebox with ppm 0.3 ppm $H_2O$ and 0.1 ppm $O_2$ levels. NPD data were measured at 10 K, 25 K 50 K, 75 K, 100 K, 125 K, 150 K, 200 K, 250 K, and 300 K (Table S5, Figs. S5, S8, S9) using neutron frame 2 with wavelength centred at 1.5 Å at the POWGEN time-of-flight diffractometer, covering a wide *d*-spacing range between 0.5 and 12 Å. The exposure time was set to around 3 h at each temperature.

**Details of Rietveld refinements**

SPXRD and NPD patterns were analysed using the software package TOPAS Academic v6.[57] First, a Pawley fit[58] of each pattern was performed to determine the lattice parameters, background, and peak shape parameters for both the main $Yb_2CsC_{60}$ and minor $Yb_2O_3$ phases. These were then carried over to the Rietveld refinements,[59,60] and the peak shape terms were fixed.

**Co-refinement against the SXRPD and NPD data sets at 100 K**: The data at 100 K were analysed via a co-refinement of the NPD and SXRPD in which all structural refining parameters were constrained to be equal between the X-ray and neutron data sets, and bond length and angle restraints were used for all C–C bonds based on an ideal $C_{60}$ molecule. The positions of all carbon and metal atoms were then refined. The occupancy of the carbon atoms comprising the fullerene units was constrained to be equal, as well as the displacement parameters, and then these were refined for all atoms. We found that the occupancy of the fullerene atoms and metal atoms refined to unity, within the experimental uncertainty, and so were fixed at unity for all subsequent refinements. The resulting co-refinement is consistent with both the X-ray and neutron data sets.

**Rietveld refinements against the variable temperature (VT) NPD patterns**: The structural model obtained at 100 K from a co-refinement was used to refine against the VT NPD data sets. Here, restraints failed to produce a model with acceptable bond lengths and angles for the fullerene anion at various temperatures, and so constraints were initially employed. In this constrained model, the atomic coordinates were fixed as those determined from the co-refinement against the 100 K data sets and then normalized with respect to the lattice parameters, effectively generating a rigid body from the 100 K data and applying it to all other temperatures. Lattice parameters were extrapolated from the SPXRD refinements, given the improved resolution, as explained below. This additional set of constraints ensured fulleride bond lengths in agreement with the literature. Given that the crystal symmetry allows for fulleride molecule rotation, we added independent rotational terms to the fractional coordinate as cos(θ), where θ was defined as 0 based on the structure determined at 100 K and allowed to deviate from zero for the refinements against the data collected at other temperatures. A further finesse was introduced by allowing for anisotropic changes to the fulleride radius, effectively producing an anisotropic expansion with *T*. Together, the final model for the fractional coordinates of the $C_{60}$ atoms that were used for the refinements against the NPD data at all temperatures is shown below.

(Eq. 1)

$$x = radius_x \times \cos(\theta_x) \times \frac{a}{a_{100\,K}} \times x_{100\,K} \ ,$$

where *x* is the fractional coordinate in $a$, $radius_x$ is the independent variable defining anisotropic expansion or compression (breathing) compared to the 100 K co-refined structure, $\cos(\theta_x)$ is the independent variable defining the rotation about *a* with respect to the 100 K co-refined structure, $\frac{a}{a_{100\,K}}$ is the lattice normalization factor for the fractional coordinates, and $x_{100\,K}$ is the fixed fractional coordinate from the co-refinement against the 100 K data set. Likewise,

(Eq. 2)

$$y = radius_y \times \cos(\theta_y) \times \frac{b}{b_{100\,K}} \times y_{100\,K} \text{ , and}$$

(Eq. 3)

$$z = radius_z \times \cos(\theta_z) \times \frac{c}{c_{100\,K}} \times z_{100\,K} \ ,$$

where the variables are as defined above. Again, the occupancies of the atoms comprising the fulleride units were fixed at 1 and the isotropic displacement parameters were constrained to be equal and this was allowed to vary. Refinements against the data using this model yielded both satisfactory fits to the patterns and chemically reasonable bond lengths and angles for the $C_{60}$ molecules at all measured temperatures.

**Rietveld refinements of the variable temperature SXRPD patterns:** Sequential Rietveld refinements against the SXRPD data sets as a function of temperature were used to extract the $Yb_2CsC_{60}$ lattice parameters, which are plotted in Fig. S7. We observed excellent agreement for the *a* and *c* lattice parameters obtained from analysis between the SXRPD and neutron data sets; however, there is slightly worse agreement for the *b* lattice parameter. To investigate this apparent disagreement, we identified the Bragg peaks in the 100 K SXRPD and NPD patterns with Miller indices, 0*k*0. Approximate positions in *Q* for these reflections are given in Table S2 based on individually Pawley fitting of either the neutron or the X-ray data. A close inspection of the (020) and the (040) Bragg peaks shows that worse resolution in the neutron data compared to the X-ray data causes these Bragg peaks (and all of the 0*k*0 reflections) to be obscured by other reflections in the neutron data, but not necessarily in the X-ray data. As such, we hypothesize that the apparent discrepancy between the *b* lattice parameter derived from neutron and X-ray data comes from a difference in resolution in these measurements, and that the synchrotron data are more reliable in determining the lattice parameters.

**Raman spectroscopy**

Raman spectra of $Yb_2CsC_{60}$ were collected on a micro-Raman LabRAM HR (HORIBA) spectrometer equipped with a Peltier-cooled charge coupled device (CCD) detector. The 632.8 nm line of a He-Ne laser was used for excitation, focused in a spot diameter of ~1 μm on the studied samples sealed in 0.5 mm-thick glass capillaries by means of a 100× objective. Laser power lower than 15 μW was used to avoid any laser-heating-induced effects, including molecular and structural modifications as well as frequency shifts of the observed Raman peaks. The elastically scattered light from the studied microcrystals was eliminated with an appropriate edge filter, while the inelastically scattered light was dispersed by using either a 600 g/mm grating for recording the Raman spectra in the whole spectral region of the intramolecular $C_{60}$ modes (spectral width: ~3.5 $cm^{-1}$) or a 1800 g/mm grating for the Raman spectra in the frequency region of the $A_g(2)$ mode of $C_{60}$ (spectral width of the system: ~0.8 $cm^{-1}$, using a confocal pinhole of 80 μm). Prior to each measurement a Ne reference lamp was employed for frequency calibration.

**X-ray absorption spectroscopy**

Partial Fluorescence Yield X-ray Absorption Spectroscopy (PFY-XAS) experiments were performed at ambient conditions at the Taiwan BL12XU beamline, SPring-8, Japan, in a 90º geometry. For the spectral acquisition, the emitted photon energy was fixed around the maximum of the $L\alpha_1$ emission line at ~7416 eV, while the incident photon energy was scanned across the Yb-$L_3$ edge.[61] Two capillaries (Ø0.3 mm, argon atmosphere) were prepared for $Yb_2CsC_{60}$ sample to confirm experimental reproducibility. In order to search for the absorption peaks corresponding to the dipolar transitions *2p* → *5d* of the 2+ and 3+ states of the ytterbium cation, PFY-XAS spectra were recorded in the energy range between 8930 and 8970 eV, divided into 120 energy steps. The data were fitted by Voigt functions to describe the spectral contributions of the dipolar excitations for divalent and trivalent Yb, superimposed on an arctangent-like function to account for the background originating from excitations toward the continuum (Fig. 2a).

**Nuclear magnetic resonance**

For nuclear magnetic resonance (NMR) experiments, approximately 44 mg of $Yb_2CsC_{60}$ powder was sealed in a custom-made Pyrex tube (outer diameter 5 mm, length 25 mm) under a partial pressure of helium of ~300 mbar. An Oxford Cryogenics continuous-flow NMR cryostat equipped with an ITC 503 temperature controller, which has a temperature stability better than ±0.1 K over the entire temperature range, was used in these experiments. $^{13}$C (nuclear spin $I$ = 1/2, natural abundance 1.109%) NMR spectra were measured for the $Yb_2CsC_{60}$ powder sample between 4 K and 340 K at a magnetic field of 9.39 T with the corresponding Larmor resonance frequency, $^{13}\nu_{ref}$ = 100.5673 MHz. As a reference, the $^{13}$C NMR signals from adamantane were used, but the shifts were then recalculated against the tetramethylsilane (TMS) standard.[62] The spectra were collected with a two-pulse Hahn echo sequence (π/2)-τ-(π)-τ and the appropriate phase cycling. The (π/2) = 4 μs length was used, while the interpulse delay was set to 100 μs. Spin-lattice relaxation times were measured with the inversion recovery technique (π)-delay-(π/2)-τ-(π)-τ where the delay time was varied from 1 ms to about $5T_1$. The $^{13}$C NMR spectra were simulated with a home-developed simulation package[14,63] to the powder axially symmetric shift anisotropy (inset to Fig. S12) as appropriate for fullerene-based solids.[14]

Pulse sequences similar to those used in $^{13}$C NMR experiments were also used to detect $^{171}$Yb (nuclear spin $I$ = ½, natural abundance 14.1%) signal at the resonance frequency of $^{171}\nu_L$ = 70.155 MHz. In this set of measurements, the (π/2) pulse length was slightly longer at 7.5 μs. For the $^{133}$Cs (nuclear spin $I$ = 7/2, natural abindance 100%) NMR experiments, we used solid-echo pulse sequence (π/2)-τ-(π/2)-τ. In measurements where the cryostat was cooled using liquid helium, pulse length (π/2) = 5 μs was used, whereas when liquid nitrogen was used, it was (π/2) = 7.5 μs. The delay τ was 60 μs.

For all three nuclei, spin-lattice relaxation times, $T_1$, reported in Fig. 2f, were extracted from the time-evolution of nuclear magnetisation $M_z(\tau)$ fitted to

(Eq. 4)

$$M_z(\tau) = M_z(0)(1 - r\exp[-(\tau/T_1)^\alpha]) \quad .$$

Here, $M_z(0)$ is the value of the nuclear (i.e., $^{13}C$, $^{171}Yb$, or $^{133}Cs$) magnetization in thermal equilibrium, $r$ is a parameter describing the degree of nuclear inversion after the π-pulse, while power-law exponent $\alpha$ is the so-called stretching exponent empirically introduced to take into account a distribution of $T_1$ values. For the studied $Yb_2CsC_{60}$ sample, $\alpha \approx 0.9$, which is a typical value for a distribution of $T_1$ in powdered fulleride samples. We note that the model described by the equation above was used to analyse the nuclear magnetisation data for the quadrupole nucleus $^{133}Cs$ too. The application of an appropriate inversion-recovery model for $I = 7/2$ resulted in a similar goodness of fit and temperature dependence of the relaxation rate.

**Density functional theory (DFT) computations.**

$Yb_2CsC_{60}$ was studied using the CASTEP plane-wave *ab initio* density functional theory (DFT) code[64] using both LDA and PBE exchange–correlation functionals[65] with ultrasoft pseudopotentials and additional Hubbard repulsion in an LSDA+$U$ scheme to include Hubbard $U$ repulsions for $C_{60}$.[66] For Yb 4$f$ and 4$d$ orbitals, an effective additional on-site Hubbard repulsions of $U_{\text{eff}} = U - J_H = 9$ eV, where $U$ is the bare Hubbard repulsion and $J_H$ is Hund's coupling, and $U_{\text{eff}} = 6.4$ eV were chosen throughout, respectively, consistently with previous LSDA+$U$ results on YbN.[67] Calculations were carried out on a DFT geometry-optimized *Pmnn* space-group (No. 58) unit cell of $Yb_2CsC_{60}$ with a 1200 eV plane-wave energy cutoff and a 4×4×3 Monkhorst-Pack grid reciprocal-space sampling[68] to achieve numerical convergence. All calculations were converged to within a tight total energy tolerance of 1 peV per atom in self-consistent field (SCF) DFT loops, while geometry-optimization calculations converged to within a 0.05 eV Å$^{-1}$ force tolerance on the atoms. Hyperfine parameters were calculated using the CASTEP NMR module,[69] while spin-resolved partial density-of-states (PDOS) band-structure DFT calculations were analysed using the OptaDOS software package.[70]

Geometry-optimization calculations indicate that the experimental structure of $Yb_2CsC_{60}$ should be very close (to within <0.015 Å, *i.e.*, close to the numerical precision of DFT) to a higher-symmetry structure described by the *Immm* space group (No. 71), which has a twice smaller primitive cell (with just one $C_{60}$ molecule per cell instead of two) and contains the initial *Pmnn* space group as one of its maximal subgroups.

## Data availability

The data that support the findings of this study are available from the corresponding author upon request. The crystallographic information files (CIFs) generated in this study have been deposited in the Cambridge Structural Database database under CCDC Deposition Numbers 2505147–2505157.

## References

1. Hubbard, J. Electron correlations in narrow energy bands. *Proc. Royal Soc. London. Series A. Mathematical and Physical Sciences* **276**, 238–257 (1997).
2. Arovas, D. P., Berg, E., Kivelson, S. A. & Raghu, S. The Hubbard Model. *Annu. Rev. Condens. Matter Phys.* **13**, 239–274 (2022).
3. Qin, M., Schäfer, T., Andergassen, S., Corboz, P. & Gull, E. The Hubbard Model: A Computational Perspective. *Annu. Rev. Condens. Matter Phys.* **13**, 275–302 (2022).
4. de' Medici, L., Mravlje, J. & Georges, A. Janus-Faced Influence of Hund's Rule Coupling in Strongly Correlated Materials. *Phys. Rev. Lett.* **107**, 256401 (2011).
5. Haddon, R. C. Electronic structure, conductivity and superconductivity of alkali metal doped ($C_{60}$). *Acc. Chem. Res.* **25**, 127–133 (1992).
6. Nomura, Y., Nakamura, K. & Arita, R. Ab initio derivation of electronic low-energy models for $C_{60}$ and aromatic compounds. *Phys. Rev. B* **85**, 155452 (2012).
7. Yue, C., Hoshino, S., Koga, A. & Werner, P. Unconventional pairing from local orbital fluctuations in strongly correlated $A_3C_{60}$. *Phys. Rev. B* **104**, 075107 (2021).
8. Capone, M., Fabrizio, M., Castellani, C. & Tosatti, E. Colloquium: Modeling the unconventional superconducting properties of expanded $A_3C_{60}$ fullerides. *Rev. Mod. Phys.* **81**, 943–958 (2009).

9. Nomura, Y., Sakai, S., Capone, M. & Arita, R. Unified understanding of superconductivity and Mott transition in alkali-doped fullerides from first principles. *Sci. Adv.* **1**, e1500568 (2015).

10. Takabayashi, Y. *et al.* The Disorder-Free Non-BCS Superconductor $Cs_3C_{60}$ Emerges from an Antiferromagnetic Insulator Parent State. *Science* **323**, 1585–1590 (2009).

11. Ganin, A. Y. *et al.* Polymorphism control of superconductivity and magnetism in $Cs_3C_{60}$ close to the Mott transition. *Nature* **466**, 221–225 (2010).

12. Ganin, A. Y. *et al.* Bulk superconductivity at 38 K in a molecular system. *Nature Mater* **7**, 367–371 (2008).

13. Klupp, G. *et al.* Dynamic Jahn–Teller effect in the parent insulating state of the molecular superconductor $Cs_3C_{60}$. *Nat Commun* **3**, 912 (2012).

14. Jeglič, P. *et al.* Low-moment antiferromagnetic ordering in triply charged cubic fullerides close to the metal-insulator transition. *Phys. Rev. B* **80**, 195424 (2009).

15. Potočnik, A. *et al.* Size and symmetry of the superconducting gap in the f.c.c. $Cs_3C_{60}$ polymorph close to the metal-Mott insulator boundary. *Sci Rep* **4**, 4265 (2014).

16. Potočnik, A. *et al.* Jahn–Teller orbital glass state in the expanded fcc $Cs_3C_{60}$ fulleride. *Chem. Sci.* **5**, 3008–3017 (2014).

17. Ihara, Y. *et al.* Spin dynamics at the Mott transition and in the metallic state of the $Cs_3C_{60}$

superconducting phases. *Europhys. Lett.* **94**, 37007 (2011).

18. Wzietek, P. *et al.* NMR Study of the Superconducting Gap Variation near the Mott Transition in $Cs_3C_{60}$. *Phys. Rev. Lett.* **112**, 066401 (2014).

19. Zadik, R. H. *et al.* Optimized unconventional superconductivity in a molecular Jahn-Teller metal. *Sci. Adv.* **1**, e1500059 (2015).

20. Rouzière, S., Margadonna, S., Prassides, K. & Fitch, A. N. Evidence of a structural anomaly at 14 K in polymerised $CsC_{60}$. *Europhys. Lett.* **51**, 314 (2000).

21. Kosaka, M. *et al.* Conducting phase of rapidly cooled $AC_{60}$ (A=Cs and Rb). *Phys. Rev. B* **51**, 12018–12021 (1995).

22. Brouet, V., Alloul, H., Quéré, F., Baumgartner, G. & Forró, L. Detection by NMR of a 'Local Spin Gap' in Quenched $CsC_{60}$. *Phys. Rev. Lett.* **82**, 2131–2134 (1999).

23. Yue, C., Hoshino, S. & Werner, P. Entropy and electronic orders of the three-orbital Hubbard model with antiferromagnetic Hund coupling. *Phys. Rev. B* **102**, 195103 (2020).

24. Yildirim, T. *et al.* Synthesis and properties of mixed alkali-metal-alkaline-earth fullerides. *Phys. Rev. B* **54**, 11981–11984 (1996).

25. Thier, K.-F. *et al.* Metallic properties of the ternary fullerides $ABa_2C_{60}$ (A= K, Rb, Cs). *Phys. Rev. B* **59**, 10536–10540 (1999).

26. Duggan, A. C. *et al.* $Ba_2CsC_{60}$: a ternary metal fulleride with a novel anion orientational

order. *Chem. Commun.* 1191–1192 (1996).

27. Tilley, T. D., Andersen, R. A., Spencer, B. & Zalkin, A. Crystal structure of bis(pentamethylcyclopentadienyl)bis(pyridine)ytterbium(II). *Inorg. Chem.* **21**, 2647–2649 (1982).

28. Shestakov, B. G. *et al.* Ytterbium(III) Complexes Coordinated by Dianionic 1,4-Diazabutadiene Ligands. *Organometallics* **34**, 1177–1185 (2015).

29. Bruno, I. J. *et al.* New software for searching the Cambridge Structural Database and visualizing crystal structures. *Acta Cryst B* **58**, 389–397 (2002).

30. Macrae, C. F. *et al.* Mercury 4.0: from visualization to analysis, design and prediction. *J Appl Cryst* **53**, 226–235 (2020).

31. Sears, V. F. Neutron scattering lengths and cross sections. *Neutron News* **3**, 26–37 (1992).

32. Momma, K. & Izumi, F. VESTA: a three-dimensional visualization system for electronic and structural analysis. *J Appl Cryst* **41**, 653–658 (2008).

33. Christides, C. *et al.* Neutron-scattering study of $C_{60}^{n-}$ ($n$=3,6) librations in alkali-metal fullerides. *Phys. Rev. B* **46**, 12088–12091 (1992).

34. Jarrige, I. *et al.* Unified understanding of the valence transition in the rare-earth monochalcogenides under pressure. *Phys. Rev. B* **87**, 115107 (2013).

35. Tanida, H., Ishihara, S., Takagi, S., Aoki, H. & Ochiai, A. $^{171}$Yb NMR studies of charge

ordering in $Yb_4As_3$. *Phys. B: Condens. Matter* **359–361**, 199–201 (2005).

36. Tycko, R. *et al.* Molecular dynamics and the phase transition in solid $C_{60}$. *Phys. Rev. Lett.* **67**, 1886–1889 (1991).

37. Rachdi, F. *et al.* High resolution $^{13}C$ NMR of $K_6C_{60}$. *Solid State Commun.* **87**, 547–550 (1993).

38. Pennington, C. H. & Stenger, V. A. Nuclear magnetic resonance of $C_{60}$ and fulleride superconductors. *Rev. Mod. Phys.* **68**, 855–910 (1996).

39. Stenger, V. A., Pennington, C. H., Buffinger, D. R. & Ziebarth, R. P. Nuclear Magnetic Resonance of $A_3C_{60}$ Superconductors. *Phys. Rev. Lett.* **74**, 1649–1652 (1995).

40. Mehring, M., Thier, K.-F., Rachdi, F. & Swiet, T. de. Localized and delocalized electronic states in $A_1C_{60}$ (A=Rb, Cs). *Carbon* **38**, 1625–1633 (2000).

41. Mitch, M. G. & Lannin, J. S. Raman scattering in $K_4C_{60}$ and $Rb_4C_{60}$ fullerenes. *Phys. Rev. B* **51**, 6784–6787 (1995).

42. Kosaka, M. *et al.* Superconductivity in $Li_xCsC_{60}$ fullerides. *Phys. Rev. B* **59**, R6628–R6630 (1999).

43. Blinc, R., Seliger, J., Dolinšek, J. & Arčon, D. Two-dimensional $^{13}C$ NMR study of orientational ordering in solid $C_{60}$. *Phys. Rev. B* **49**, 4993–5002 (1994).

44. Gosar, Ž. & Arčon, D. Determination of symmetry of a superconducting order parameter

in nuclear magnetic resonance experiments. *Contemp. Phys.* **65**, 163–179 (2024).

45. Walstedt, R. E. *The NMR Probe of High-Tc Materials and Correlated Electron Systems*. vol. 276 (Springer, Berlin, Heidelberg, 2018).

46. Tanigaki, K. *et al.* Influence of rotational disorder in $C_{60}$ on electrical conductivity. *J. Phys. Chem. Solids* **181**, 111537 (2023).

47. Arčon, D., Prassides, K., Maniero, A.-L. & Brunel, L.C. High-field electron paramagnetic resonance study of the polymerization in $Na_2Rb_{1-x}Cs_xC_{60}$. *Appl. Magn. Res.* **19**, 531 (2000).

48. Brouet, V., Alloul, H., Garaj, S. & Forró, L. Persistence of molecular excitations in metallic fullerides and their role in a possible metal to insulator transition at high temperatures. *Phys. Rev. B* **66**, 155124 (2002).

49. Özdaş, E. *et al.* Superconductivity and cation-vacancy ordering in the rare-earth fulleride $Yb_{2.75}C_{60}$. *Nature* **375**, 126–129 (1995).

50. Margadonna, S., Arvanitidis, J., Papagelis, K. & Prassides, K. Negative Thermal Expansion in the Mixed Valence Ytterbium Fulleride, $Yb_{2.75}C_{60}$. *Chem. Mater.* **17**, 4474–4478 (2005).

51. Klupp, G. & Kamarás, K. Following Jahn–Teller Distortions in Fulleride Salts by Optical Spectroscopy. in *The Jahn-Teller Effect: Fundamentals and Implications for Physics and*

*Chemistry* (eds Köppel, H., Yarkony, D. R. & Barentzen, H.) 489–515 (Springer, Berlin, Heidelberg, 2009).

52. Chancey, C. C. & O'Brien, M. C. M. *The Jahn-Teller Effect in $C_{60}$ and Other Icosahedral Complexes*. (Princeton University Press, 1997).

53. Potočnik, A., Manini, N., Komelj, M., Tosatti, E. & Arčon, D. Orthorhombic fulleride $(CH_3NH_2)K_3C_{60}$ close to Mott-Hubbard instability: Ab initio study. *Phys. Rev. B* **86**, 085109 (2012).

54. Koch, E., Gunnarsson, O. & Martin, R. M. 1999. Screening, Coulomb pseudopotential, and superconductivity in alkali-doped fullerenes. *Phys. Rev. Lett.* **83**, 620–623 (1999).

55. Isidori, A. *et al.* Charge Disproportionation, Mixed Valence, and Janus Effect in Multiorbital Systems: A Tale of Two Insulators. *Phys. Rev. Lett.* **122**, 186401 (2019).

56. Huq, A. *et al.* POWGEN: rebuild of a third-generation powder diffractometer at the Spallation Neutron Source. *J Appl Cryst* **52**, 1189–1201 (2019).

57. Coelho, A. A. TOPAS and TOPAS-Academic: an optimization program integrating computer algebra and crystallographic objects written in C++. *J Appl Cryst* **51**, 210–218 (2018).

58. Pawley, G. S. Unit-cell refinement from powder diffraction scans. *J Appl Cryst* **14**, 357–361 (1981).

59. Von Dreele, R. B., Jorgensen, J. D. & Windsor, C. G. Rietveld refinement with spallation neutron powder diffraction data. *J Appl Cryst* **15**, 581–589 (1982).

60. Rietveld, H. M. A profile refinement method for nuclear and magnetic structures. *J Appl Cryst* **2**, 65–71 (1969).

61. Yamaoka, H. *et al.* Electronic structure of $YbGa_{1.15}Si_{0.85}$ and $YbGa_xGe_{2-x}$ probed by resonant x-ray emission and photoelectron spectroscopies. *Phys. Rev. B* **83**, 104525 (2011).

62. Hoffman, R. Solid-state chemical-shift referencing with adamantane. *J. Magn. Reson.* **340**, 107231 (2022).

63. Potočnik, A. *et al.* Anomalous local spin susceptibilities in noncentrosymmetric $La_2C_3$ superconductor. *Phys. Rev. B* **90**, 104507 (2014).

64. Clark, S. J. *et al.* First principles methods using CASTEP. *Z. fur Krist. - Cryst. Mater.* **220**, 567–570 (2005).

65. Perdew, J. P., Burke, K. & Ernzerhof, M. Generalized Gradient Approximation Made Simple. *Phys. Rev. Lett.* **77**, 3865–3868 (1996).

66. Anisimov, V. I., Aryasetiawan, F. & Lichtenstein, A. I. First-principles calculations of the electronic structure and spectra of strongly correlated systems: the LDA+ U method. *J. Phys.: Condens. Matter* **9**, 767 (1997).

67. Larson, P., Lambrecht, W. R. L., Chantis, A. & van Schilfgaarde, M. Electronic structure of rare-earth nitrides using the LSDA+U approach: Importance of allowing 4f orbitals to break the cubic crystal symmetry. *Phys. Rev. B* **75**, 045114 (2007).

68. Monkhorst, H. J. & Pack, J. D. Special points for Brillouin-zone integrations. *Phys. Rev. B* **13**, 5188–5192 (1976).

69. Bonhomme, C. *et al.* First-Principles Calculation of NMR Parameters Using the Gauge Including Projector Augmented Wave Method: A Chemist's Point of View. *Chem. Rev.* **112**, 5733–5779 (2012).

70. Nicholls, R. J., Morris, A. J., Pickard, C. J. & Yates, J. R. OptaDOS - a new tool for EELS calculations. *J. Phys.: Conf. Ser.* **371**, 012062 (2012).

## Acknowledgements

We thank SPring-8 for access to synchrotron X-ray facilities (Proposals no. 2021A1587, 2021A4254, 2022B4252, and 2023B2101). Neutron diffraction measurements were performed on the POWGEN diffractometer at the Spallation Neutron Source, a DOE Office of Science User Facility operated by the Oak Ridge National Laboratory (Proposal no. IPTS-29846). We acknowledge fruitful discussions with Y. Takabayashi and M. Capone, and the assistance of A. Gregorovič and A. Rešetič during the high-resolution $^{13}$C NMR measurements.

## Funding statement

K.P. discloses support for the research of this work from JSPS KAKENHI Grants-in-Aid for Scientific Research (grant numbers JP21H01907, JP22K18693, and JP23KJ1843), N.Y. from JST SPRING (grant number JPMJSP2138), K.M. from JST SPRING (grant number JPMJSP2139), A.D. from the Slovenian Research Agency (grant numbers P1-0125, J1-3007, and J1-70050), and M.G. from the Slovenian Research Agency (grant numbers J1-50012 and N1-0356). C.M.B. discloses full support and R.A.K. partial support for the research of this work from NIST.

## Author contributions

D.A., C.M.B. and K.P. directed and coordinated the project. K.M. synthesized the samples. K.M., N.Y. and S.K. performed the synchrotron XRPD measurements, and C.M.B., K.P., and Q.Z. performed the neutron PD measurements. R.A.K. and C.M.B. analysed the diffraction data in close contact with K.P. following preliminary work by K.M. and N.Y. These results were discussed with S.M., H.Ishibashi and Y.K. Raman measurements were performed by J.A. who also analysed the data. K.M., N.Y., J.A., H.Y., N.H. and H.Ishii carried out the PFY-XAS measurements and K.M. analysed the data. D.A. and U.K. performed the NMR measurements and D.A. analysed the data. M.G. carried out electronic structure calculations and interpreted the results. D.A., C.M.B. and K.P. wrote the paper. All authors commented on the paper.

### Competing Interest Statement

The authors declare no competing interests.

### Main text figure captions

**Figure 1 │ Crystal structure of $Yb_2CsC_{60}$. a,** Schematic illustration of the structural model of the orthorhombic crystal structure of $Yb_2CsC_{60}$ (unit cell depicted by red $C_{60}^{5-}$ anions) and its relationship with the parent cubic-close–packed structures of ternary $A_2A'C_{60}$ fullerides ($A$, $A'$ = alkali metals) and its $Ba_2CsC_{60}$ analogue (unit cell depicted by blue $C_{60}^{3-}$ anions in only one of the two possible merohedrally disordered orientations) as described in the text. The $Yb^{2+}$ and $Cs^{+}$ ions are shown as green and purple spheres, respectively. **b,** Local packing of the $Cs^{+}$ cation (purple sphere) in the octahedral void with four hexagon-pentagon (6:5) C–C bonds of the surrounding $C_{60}$ units in the *ab* plane and two hexagon-hexagon (6:6) C–C bonds of the surrounding $C_{60}$ units along the *c* axis direction. **c,** Local packing of the $Yb^{2+}$ cation (green sphere) in the tetrahedral site of the orthorhombic unit cell. The $Yb^{2+}$ cation faces two pentagonal faces and two hexagon-pentagon (6:5) C–C bonds of the surrounding $C_{60}$ units. **d,** Relative change in the orthorhombic lattice parameters with increasing temperature normalized to the values derived from the Rietveld refinement against the NPD data collected at 10 K, illustrating the anisotropic expansion along *a*. Change **e,** in the *y* fractional coordinate of the $Yb^{2+}$ cation and **f,** in the distance between the Yb cation and the centroids of the pentagonal faces of near-neighbour $C_{60}$ anions along the <011> direction as a function of temperature. **g,** Ratio of isotropic displacement parameter, $b_{eq}$, divided by their 10 K values, $b_{eq}^{10\,K}$, for Yb (circles), Cs (triangles), and C (rhombi) as a function of temperature. Unless shown, error bars are smaller than or commensurate with symbols and represent an uncertainty of $\pm1\sigma$.

**Figure 2 │ Electronic ground state of $Yb_2CsC_{60}$. a,** Yb-$L_3$ PFY-XAS spectrum of $Yb_2CsC_{60}$ powder measured at room temperature. Green open circles, thick grey solid line, and black dashed line correspond to experimental data, total calculated intensity, and the background intensity, respectively. Total intensity of 2+ (3+) ytterbium oxidation states is shown as a green (blue) filled area. **b,** $^{171}$Yb NMR spectra of $Yb_2CsC_{60}$ powder at selected temperatures between 290 and 7 K. The reference frequency in these measurements was 70.155 MHz. The single peak in the $^{171}$Yb NMR spectra corresponds to $Yb^{2+}$ cations only, without any discernible component that could be attributed to $Yb^{3+}$ sites. **c,** $^{13}$C magic angle spinning (MAS) NMR spectrum of $Yb_2CsC_{60}$ powder measured at room temperature. The shift of 168 ppm is measured relative to TMS standard. The spectrum has been taken at a resonance frequency of 125.768304 MHz and at a spinning frequency of 25 kHz. The chemical shifts for neutral $C_{60}$ and hexavalent $C_{60}^{6-}$ are 143 ppm (vertical dashed red line) and 156 ppm (vertical dotted blue line), respectively. **d,** Typical Raman spectrum of $Yb_2CsC_{60}$ samples expanded in the frequency region of the $A_g(2)$ vibrational mode of $C_{60}$. The $A_g(2)$ mode is redshifted to 1443 cm$^{-1}$ (vertical dashed purple line) from the 1469 cm$^{-1}$ (vertical dashed red line) characteristic of neutral $C_{60}$. **e,** Temperature dependence of NMR shifts for $^{13}$C (grey circles, left scale), $^{171}$Yb (green circles, right scale) and $^{133}$Cs (purple circles, right scale) nuclei determined from the fits of the corresponding NMR spectra (see Methods). **f,** Temperature dependence of the inverse $^{13}$C spin lattice relaxation time, $^{13}T_1$ divided by temperature (grey circles, left scale). Horizontal dashed line at $1/^{13}T_1T = 0.93\times10^{-3}$ K$^{-1}$s$^{-1}$ marks the Korringa relaxation contribution dominant at low temperatures. Temperature dependences of the inverse $^{171}$Yb spin lattice relaxation time, $^{171}T_1$ (green circles, right scale) and $^{133}$Cs spin lattice relaxation time, $^{133}T_1$ (purple circles, right scale) divided by temperature are shown for comparison. Error bars represent an uncertainty of $\pm1\sigma$.

**Figure 3. Electronic band structure of orthorhombic $Yb_2CsC_{60}$. a.** DFT calculated band dispersions for $Yb_2CsC_{60}$ in the structurally relaxed *Immm* space group. The unit cell contains two $C_{60}$ cages and thus a total of six $t_{1u}$-derived bands are present close to the Fermi energy, $E_F$ (dashed horizontal line). The energies are plotted along lines in the Brillouin zone connecting points Γ = (0,0,0), X = (½,0,0), S = (½,½,0), Y = (0,½,0), Z = (0,0,½) and R = (½,½,½). These "$C_{60}$ LUMO ($t_{1u}$)" bands are clearly separated from the "$C_{60}$ LUMO+1 ($t_{1g}$)" ones, which are found at higher energy by ~0.3 eV and the "$C_{60}$ HOMO ($h_u$)" ones, which are lower in energy by ~0.4 eV lower. The *right* panel shows the calculated total density-of-states, $g(E)$, which yields $g(E_F)$ = 5.4 states/eV/$C_{60}$ arising almost exclusively from the $t_{1u}$-derived bands. The Yb *d*- and *f*- orbital contributions to the total density-of-states are 1.56% and 0.28%, respectively and are thus negligible within the precision of the DFT computations. **b.** Schematic representation of the effect of the orthorhombic crystal-field, $\Delta_{cf}$, on the single-hole $t_{1u}$ systems. The two $C_{60}^{5-}$ anions per unit cell in $Yb_2CsC_{60}$ contribute with their $t_{1u}$ orbitals (left and right columns of the molecular electronic levels) in total six $t_{1u}$ orbitals to the conduction band. In the limit of large $\Delta_{cf}$, a single half-filled band is separated from the other two: in this case in the absence of effective Hund's exchange a Mott-insulating instability with an antiferromagnetic exchange, $J$, between the two unpaired spins of $C_{60}^{5-}$ anions is expected due to electron correlations (top panel). However, in the limit of moderate $\Delta_{cf}$, partial spilling of states between the sub-bands will recover the influence of an effective Hund's exchange, $J_{eff}$, and assist in stabilizing a metallic state (bottom panel).

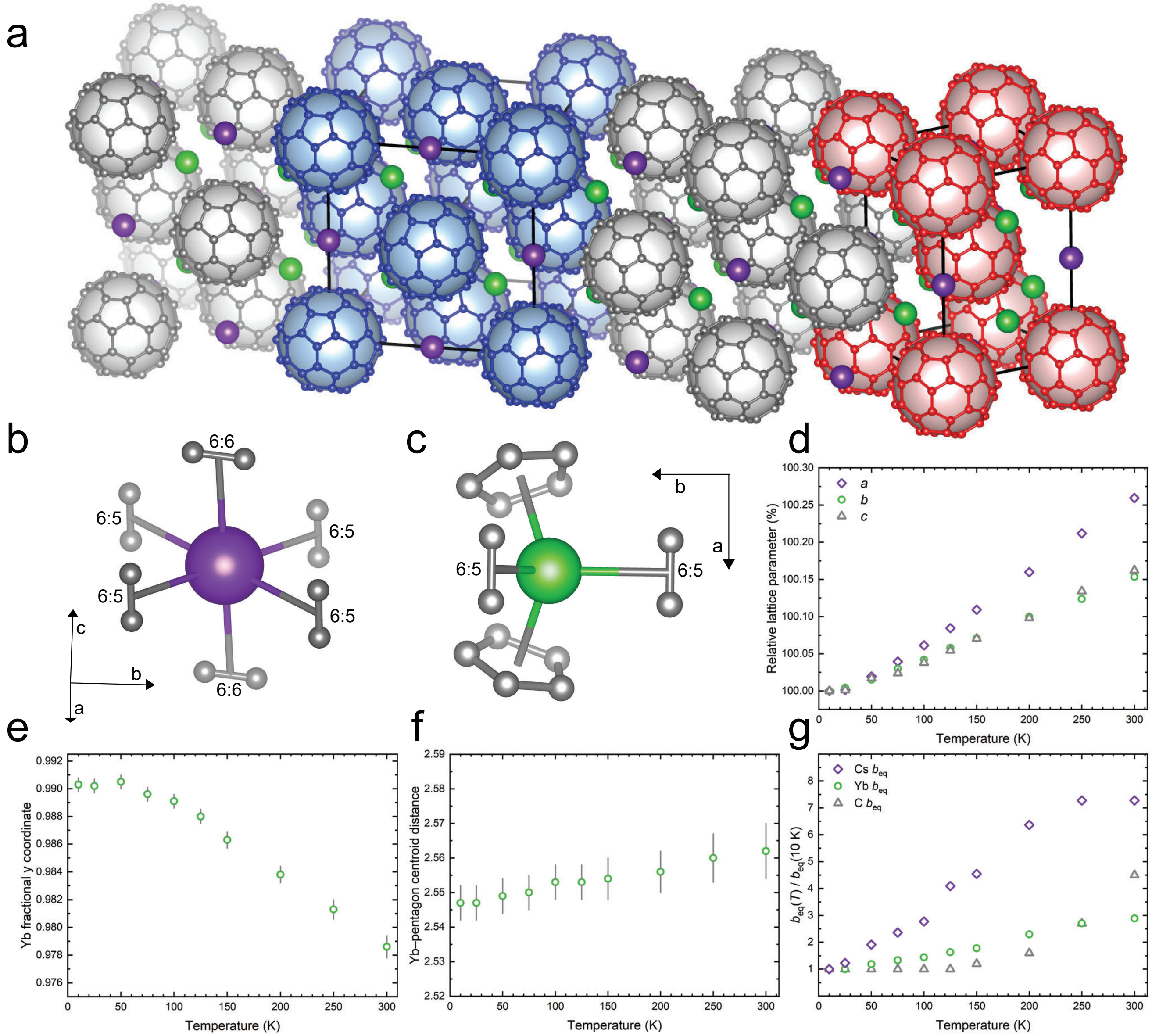

a
b
6:6
6:5
6:5
6:5
6:5
6:6
c
b
a
c
b
a
6:5
6:5
d
Relative lattice parameter (%)
a
b
c
Temperature (K)
e
Yb fractional y coordinate
Temperature (K)
f
Yb–pentagon centroid distance
Temperature (K)
g
Cs $b_{eq}$
Yb $b_{eq}$
C $b_{eq}$
$b_{eq}(T) / b_{eq}(10 K)$
Temperature (K)

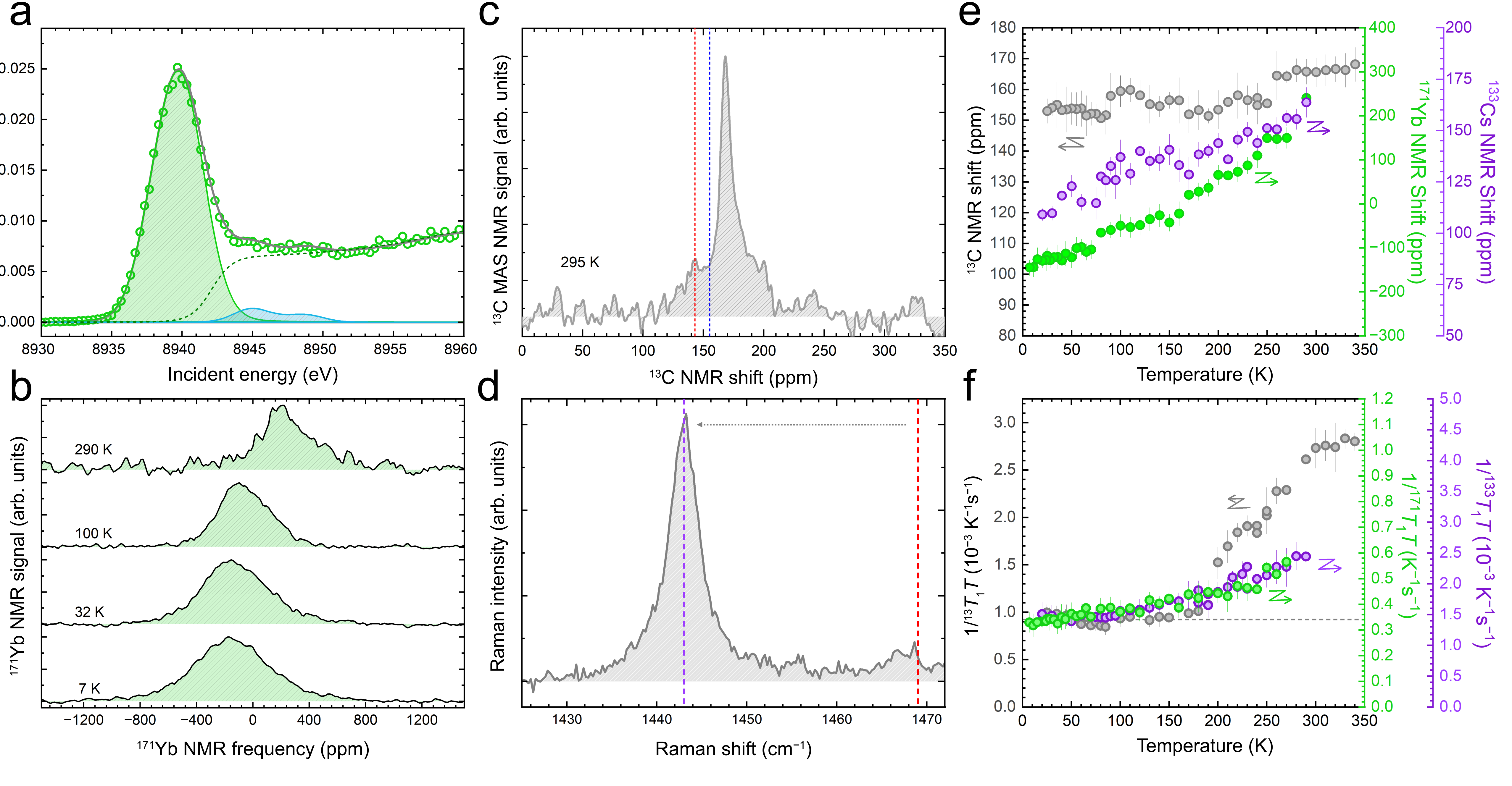

a
0.025
0.020
0.015
0.010
0.005
0.000
8930
8935
8940
8945
8950
8955
8960
Incident energy (eV)
b
290 K
100 K
32 K
7 K
171Yb NMR signal (arb. units)
−1200
−800
−400
0
400
800
1200
171Yb NMR frequency (ppm)
c
295 K
13C MAS NMR signal (arb. units)
0
50
100
150
200
250
300
350
13C NMR shift (ppm)
d
Raman intensity (arb. units)
1430
1440
1450
1460
1470
Raman shift (cm−1)
e
13C NMR shift (ppm)
171Yb NMR Shift (ppm)
133Cs NMR Shift (ppm)
Temperature (K)
f
1/13T1T (10−3 K−1s−1)
1/171T1T (K−1s−1)
1/133T1T (10−3 K−1s−1)
Temperature (K)

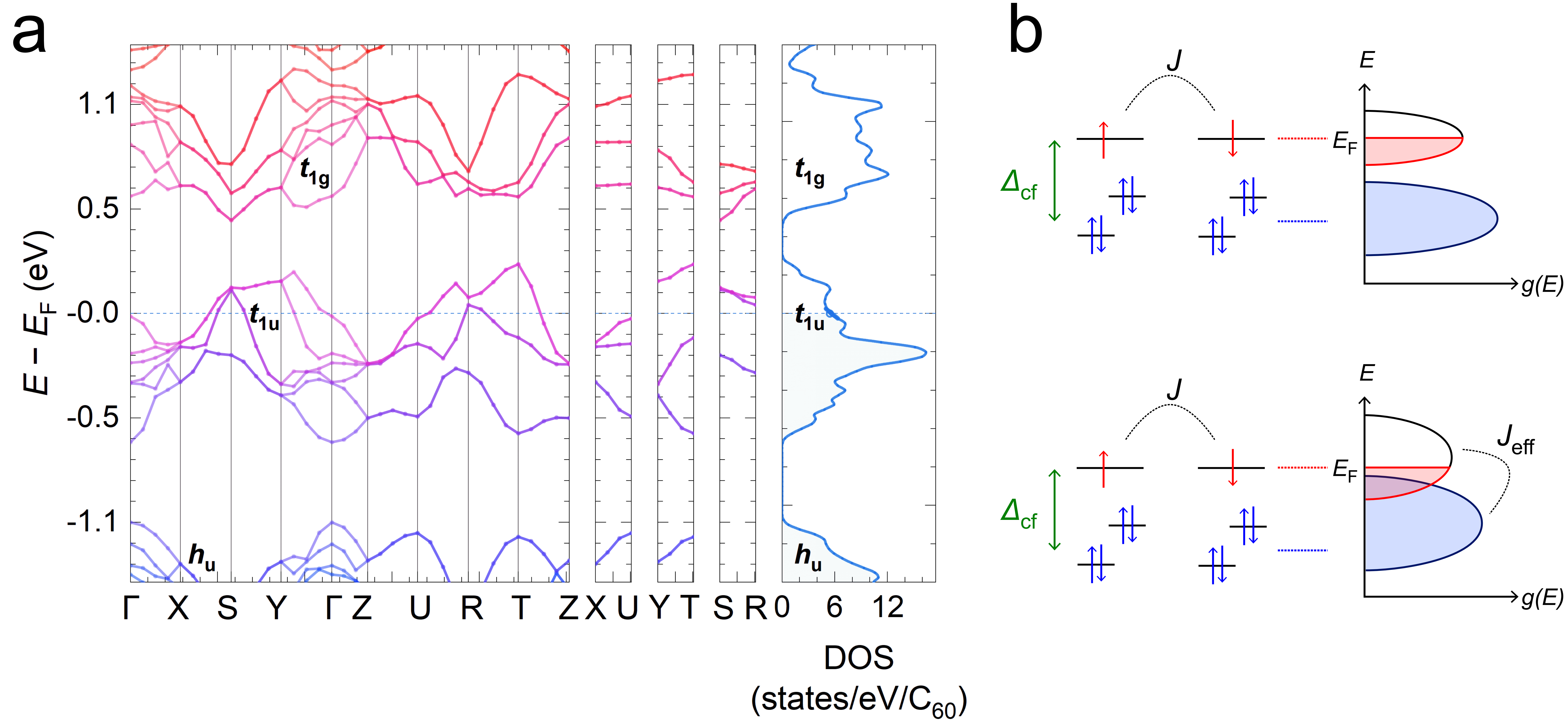

a
b
1.1
0.5
-0.0
-0.5
-1.1
E - E_F (eV)
t_1g
t_1u
h_u
Γ X S Y Γ Z U R T Z X U Y T S R
0 6 12
DOS
(states/eV/C60)
J
E
E_F
Δcf
g(E)
J_eff

# Supplementary information

# Survival of the metallic state in a single-hole multiband *p*-orbital molecular system

Keisuke Matsui[1#], Ryan A. Klein[2,3#], Naoya Yoshikane[4], John Arvanitidis[5], Matjaž Gomilšek[6], Urh Klopčič[6], Shogo Kawaguchi[7], Hitoshi Yamaoka[8], Nozomu Hiraoka[9], Hirofumi Ishii[9], Qiang Zhang[10], Shigeo Mori[1], Hiroki Ishibashi[4], Yoshiki Kubota[4], Craig M. Brown[3*], Denis Arčon[6,11*] & Kosmas Prassides[4,6*]

[1] Department of Materials Science, Graduate School of Engineering, Osaka Metropolitan University, Osaka 599-8531, Japan

[2] Department of Chemistry, University of California, Berkeley, California 94720, United States

[3] Center for Neutron Research, National Institute of Standards and Technology (NIST), Gaithersburg, Maryland 20899, USA

[4] Department of Physics, Graduate School of Science, Osaka Metropolitan University, Osaka 558-8585, Japan

[5] Physics Department, Aristotle University of Thessaloniki, Thessaloniki 54124, Greece

[6] Institute "Jožef Stefan", Jamova 39, SI-1000 Ljubljana, Slovenia

[7] Japan Synchrotron Radiation Research Institute (JASRI), 1-1-1 Kouto, Sayo-cho, Sayo-gun, Hyogo 679-5198, Japan

[8] RIKEN SPring-8 Center, Sayo, Hyogo 679-5148, Japan

[9] National Synchrotron Radiation Research Center, Hsinchu 30076, Taiwan

[10] Neutron Scattering Division, Oak Ridge National Laboratory, Oak Ridge, Tennessee 37831, USA

[11] Faculty of Mathematics and Physics, University of Ljubljana, Jadranska 19, SI-1000 Ljubljana, Slovenia

[#] These authors contributed equally to the manuscript.

* e-mail: craig.brown@nist.gov; denis.arcon@ijs.si; kosmas.prassides@ijs.si

**The relationship between the NMR spin-lattice relaxation rate, $1/T_1$ and the isotropic nuclear displacement parameter, $b_{eq}$**

In paramagnetic systems, one must take into account the hyperfine coupling between the nuclear spin **I** and the electronic spins **S**, i.e., the corresponding spin Hamiltonian can be written as $H_{hf} = \mathbf{I}\cdot\mathbf{A}\cdot\mathbf{S}$. Here, **A** is the hyperfine coupling tensor, which is a second-rank tensor comprising contributions from the isotropic hyperfine coupling and the anisotropic (i.e. dipolar) coupling, respectively. Fluctuations of the electronic spins make $H_{hf}$ time-dependent and thus determine the nuclear magnetic resonance (NMR) spin–lattice relaxation rate, $1/T_1$, which can be expressed as follows:[S1]

$$\frac{1}{T_1 T} = \frac{2\gamma_n^2 k_B}{(g\mu_B)^2} \sum_{\boldsymbol{q}} \boldsymbol{A}_{\boldsymbol{q}} \boldsymbol{A}_{-\boldsymbol{q}} \frac{\chi''(\boldsymbol{q}, \omega)}{\omega}$$

where $\gamma_n$ is the nuclear gyromagnetic ratio, $\mu_B$ is the Bohr magneton, and $\chi''(\boldsymbol{q}, \omega)$ denotes the imaginary part of the dynamic magnetic spin susceptibility. The Korringa relation used to discuss results in the main text can be directly derived from the above expression – see for example Ref. S1. Due to nuclear displacements, $u_n$, the components of the hyperfine coupling tensor are slightly modified. Expanding the components of $A$, one obtains

$$A_{\boldsymbol{q}}(u_n) = A_{\boldsymbol{q}}(0) + \frac{\partial A_{\boldsymbol{q}}}{\partial u_n} u_n + \cdots = A_{\boldsymbol{q}}(0) + a_{n,\boldsymbol{q}} u_n + \cdots.$$

The constant $a_{n,\boldsymbol{q}}$ quantifies the change in the hyperfine coupling constant with respect to the nuclear displacement. Since $1/T_1T$ is proportional to the square of the hyperfine coupling constant, it is clear that $\langle A_{\boldsymbol{q}} A_{-\boldsymbol{q}} \rangle \rightarrow A_{\boldsymbol{q}}(0) A_{-\boldsymbol{q}}(0) + a_{n,\boldsymbol{q}} a_{n,-\boldsymbol{q}} \langle u_n^2 \rangle$, since $\langle u_n \rangle = 0$, and $\langle u_n^2 \rangle$ is the mean-square nuclear displacement averaged over time and directions. Inserting this into the above expression for $1/T_1T$, we thus establish a direct link between the variation of $1/T_1T$ and the isotropic nuclear displacement parameter, $b_{eq} = \frac{8\pi^2}{3} \langle u_n^2 \rangle$.

## Reference

S1. Gosar, Ž. & Arčon, D. Determination of symmetry of a superconducting order parameter in nuclear magnetic resonance experiments. *Contemp. Phys.* **65**, 163–179 (2024).

**Table S1** | Refined parameters for $Yb_2CsC_{60}$ (space group *Pmnn*) obtained from the combined Rietveld analysis of synchrotron X-ray (SXRPD) and neutron (NPD) powder diffraction data collected at 100 K. Estimated uncertainties in the last digits are given in parentheses and represent ±1σ (i.e., 1 standard deviation). The fit statistics for the co-refinement are $R_{wp}$ = 4.720 %, $R_p$ = 3.437 %, $R_{exp}$ = 1.902 %, GOF = 2.482. The lattice constants and unit cell volume for the main phase are $a$ = 9.97392(15) Å, $b$ = 9.76416(16) Å, and $c$ = 13.8828(2) Å, and $V$ = 1352.01(4) Å$^3$, respectively. The fractions of the co-existing cubic $Yb_2O_3$ impurity phase (space group $Ia\bar{3}$, $a$ = 10.429(2) Å) are 7.5(1) % in the SXRPD and 2.3(1) % in the NPD data, respectively.

| atom | *x/a* | *y/b* | *z/c* | Occupancy | $b_{eq}$ (Å$^2$) |
|---|---|---|---|---|---|
| C1 | 0.1163(5) | 0.1258(4) | 0.2243(4) | 1 | 0.010(6) |
| C2 | 0.2323(5) | −0.0896(4) | 0.1838(3) | 1 | 0.010(6) |
| C3 | 0.2292(5) | 0.0586(4) | 0.1893(3) | 1 | 0.010(6) |
| C4 | 0.1175(5) | −0.1630(4) | 0.2111(3) | 1 | 0.010(6) |
| C5 | 0.0736(6) | −0.2798(5) | 0.1555(3) | 1 | 0.010(6) |
| C6 | 0.1441(5) | −0.3174(4) | 0.0719(3) | 1 | 0.010(6) |
| C7 | 0.3058(5) | −0.1292(5) | 0.0981(3) | 1 | 0.010(6) |
| C8 | 0.2599(5) | −0.2398(4) | 0.0427(3) | 1 | 0.010(6) |
| C9 | 0.3003(5) | 0.1109(4) | 0.1062(3) | 1 | 0.010(6) |
| C10 | 0.3479(3) | −0.0042(5) | 0.0505(3) | 1 | 0.010(6) |
| C11 | 0.0698(3) | −0.3558(4) | −0.0147(3) | 1 | 0.010(6) |
| C12 | 0.2594(6) | 0.2321(4) | 0.0612(3) | 1 | 0.010(6) |
| C13 | 0.0733(5) | 0.2521(4) | 0.1782(4) | 1 | 0.010(6) |
| C14 | 0.1423(5) | 0.3031(5) | 0.0977(3) | 1 | 0.010(6) |
| C15 | 0 | −0.0948(6) | 0.2497(5) | 1 | 0.010(6) |
| C16 | 0 | 0.0485(5) | 0.2568(5) | 1 | 0.010(6) |
| Cs | 0 | 0 | 0.5 | 1 | 0.52(4) |
| Yb | 0.5 | 0.99162(18) | 0.22422(5) | 1 | 0.308(18) |

**Table S2 │** Miller indices for $0k0$ type reflections derived from Pawley fitting the X-ray (SXRPD) and neutron (NPD) data separately.

| $h$ | $k$ | $l$ | $Q$ (Å$^{-1}$) SXRPD | $Q$ (Å$^{-1}$) NPD |
|---|---|---|---|---|
| 0 | 2 | 0 | 1.2871 | 1.2856 |
| 0 | 4 | 0 | 2.5743 | 2.5711 |
| 0 | 6 | 0 | 3.8614 | 3.8567 |
| 0 | 8 | 0 | 5.1485 | 5.1423 |
| 0 | 10 | 0 | 6.4356 | 6.4279 |
| 0 | 12 | 0 | 7.7228 | 7.7134 |
| 0 | 14 | 0 | 9.0100 | 8.9990 |
| 0 | 16 | 0 | - | 10.2846 |

**Table S3:** First principles calculations of the isotropic hyperfine coupling ($a_{iso}$), chemical shift ($\sigma$) and electric field gradients (EFG) for the two different $Yb_2CsC_{60}$ structures: experimental *Pmnn* structure and the DFT geometry-optimised *Immm* structure.

| Nucleus | $a_{iso}$ (kHz) | $\sigma$ (ppm) | EFG (MHz) |
|---|---|---|---|
| *Pmnn* structure (experiment) | | | |
| Yb | 0.0038 | 9474.5 | 366.1 |
| Cs | −0.0092 | 5767.3 | 0.109 |
| *Immm* structure (DFT geometry optimisation) | | | |
| Yb | 1.706 | 6317 | 529.9 |
| Cs | −1.296 | 2430.6 | 0.103 |

**Table S4 │** Fulleride diametric distances and ratios, as illustrated in Fig. S6. Values in parentheses represent an experimental uncertainty of ±1 standard deviation σ.

| *T* (K) | cross *a* $C_{10}$–$C_{10}$ (Å) | cross *b* $C_{11}$–$C_{11}$ (Å) | cross *c* $C_{15}$–$C_{16}$ (Å) | $R_c/R_a$ | $R_c/R_b$ |
|---|---|---|---|---|---|
| 10 | 6.957(3) | 6.965(3) | 7.0472(16) | 1.0130(5) | 1.0118(5) |
| 25 | 6.957(3) | 6.965(3) | 7.0462(17) | 1.0128(5) | 1.0117(5) |
| 50 | 6.957(3) | 6.966(3) | 7.0454(18) | 1.0127(5) | 1.0114(5) |
| 75 | 6.958(3) | 6.962(3) | 7.0443(18) | 1.0124(5) | 1.0118(5) |
| 100 | 6.957(3) | 6.961(3) | 7.045(18) | 1.0127(5) | 1.0121(5) |
| 125 | 6.957(3) | 6.964(3) | 7.047(2) | 1.0129(5) | 1.0120(5) |
| 150 | 6.959(3) | 6.963(3) | 7.048(2) | 1.0128(5) | 1.0122(5) |
| 200 | 6.957(3) | 6.964(4) | 7.049(3) | 1.0132(6) | 1.0122(7) |
| 250 | 6.955(4) | 6.965(4) | 7.052(3) | 1.0140(7) | 1.0125(7) |
| 300 | 6.958(4) | 6.966(4) | 7.053(3) | 1.0140(7) | 1.0125(7) |

**Table S5 │** Lattice parameters *a*, *b*, and *c*, the derived unit cell volume *V*, and the $R_{wp}$ fit statistic from sequential Rietveld refinements of the NPD data as a function of temperature. Values in parentheses indicate ±1σ.

| $T$ (K) | $a$ (Å) | $b$ (Å) | $c$ (Å) | $V$ (Å$^3$) | $R_{wp}$ (%) |
|---|---|---|---|---|---|
| 10 | 9.9678(3) | 9.7601(4) | 13.8775(5) | 1350.10(8) | 4.60399 |
| 25 | 9.9679(3) | 9.7605(4) | 13.8777(5) | 1350.19(8) | 4.74526 |
| 50 | 9.9697(3) | 9.7615(4) | 13.8799(5) | 1350.80(9) | 4.82178 |
| 75 | 9.9717(3) | 9.7630(4) | 13.8809(5) | 1351.36(9) | 4.70039 |
| 100 | 9.9739(3) | 9.7642(4) | 13.8828(6) | 1352.01(9) | 4.79328 |
| 125 | 9.9762(4) | 9.7657(4) | 13.8851(6) | 1352.75(10) | 4.8381 |
| 150 | 9.9787(4) | 9.7670(5) | 13.8873(6) | 1353.48(10) | 5.01571 |
| 200 | 9.9837(4) | 9.7698(5) | 13.8912(7) | 1354.93(12) | 5.20271 |
| 250 | 9.9889(5) | 9.7721(6) | 13.8962(8) | 1356.44(13) | 5.45834 |
| 300 | 9.9937(6) | 9.7750(7) | 13.9001(10) | 1357.88(16) | 5.85637 |

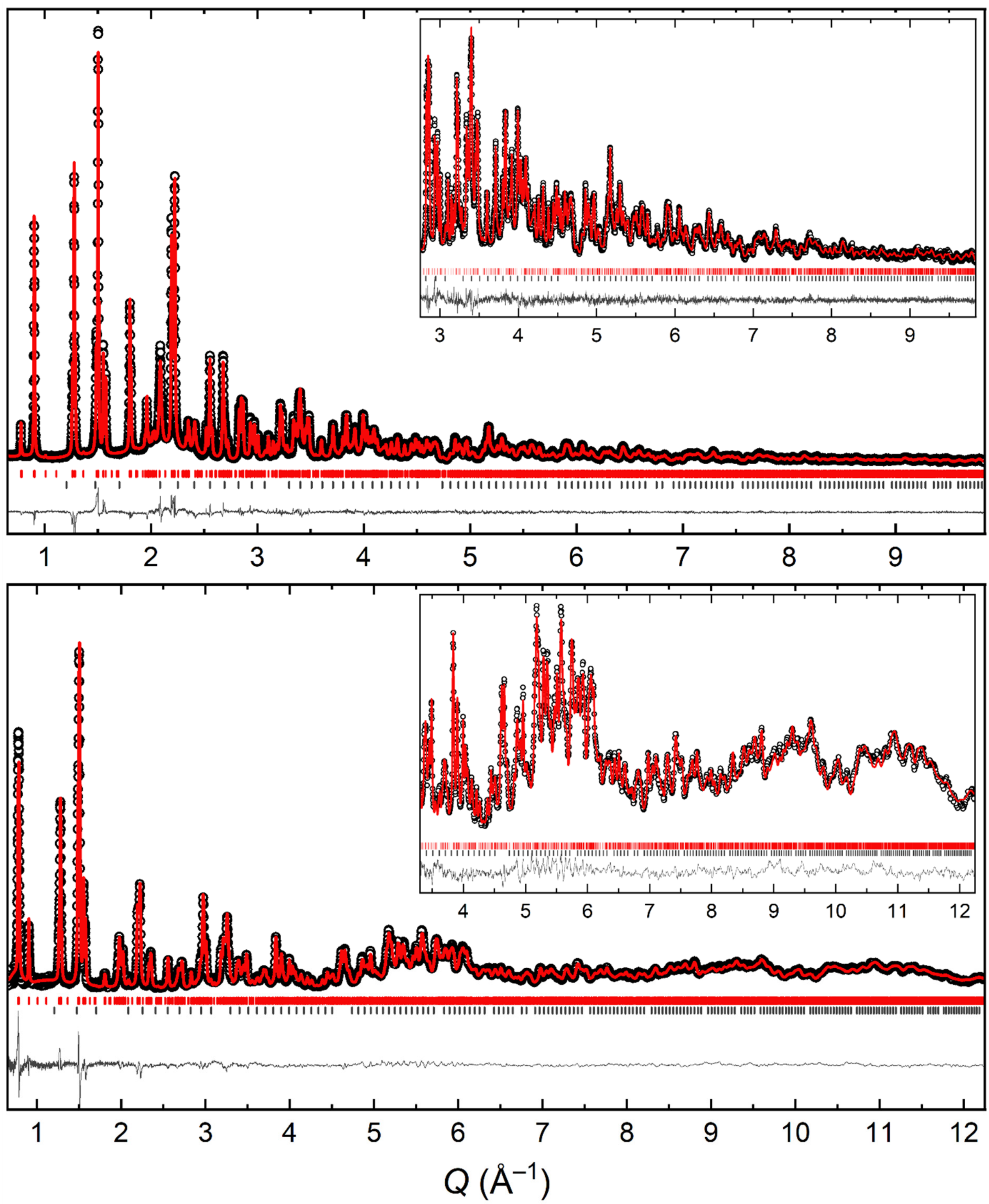

**Figure S1 │ Structural characterisation of $Yb_2CsC_{60}$.** Simultaneous Rietveld co-refinements of the SXRPD (top) and NPD (bottom) patterns for $Yb_2CsC_{60}$ (space group *Pmnn;* $a$ = 9.97392(15) Å, $b$ = 9.76416(16) Å, $c$ = 13.8828(2) Å, and $V$ = 1352.01(4) Å$^3$) obtained at 100 K. Black circles depict the data while red and grey traces are the Rietveld and difference curves, respectively. The top set of red vertical tick marks indicates the *hkl* positions for $Yb_2CsC_{60,}$ while the bottom set of black vertical tick marks indicates *hkl* positions from the minor $Yb_2O_3$ impurity phase. $R_{wp}$ = 4.720%, $R_p$ = 3.437%, $R_{\mathrm{exp}}$ = 1.902%, GOF = 2.482. The resulting structural parameters are tabulated in Table S1. Symbols are larger than or commensurate with their error bars, which represent ±1σ. The insets illustrate the quality of the fits at high scattering $Q$.

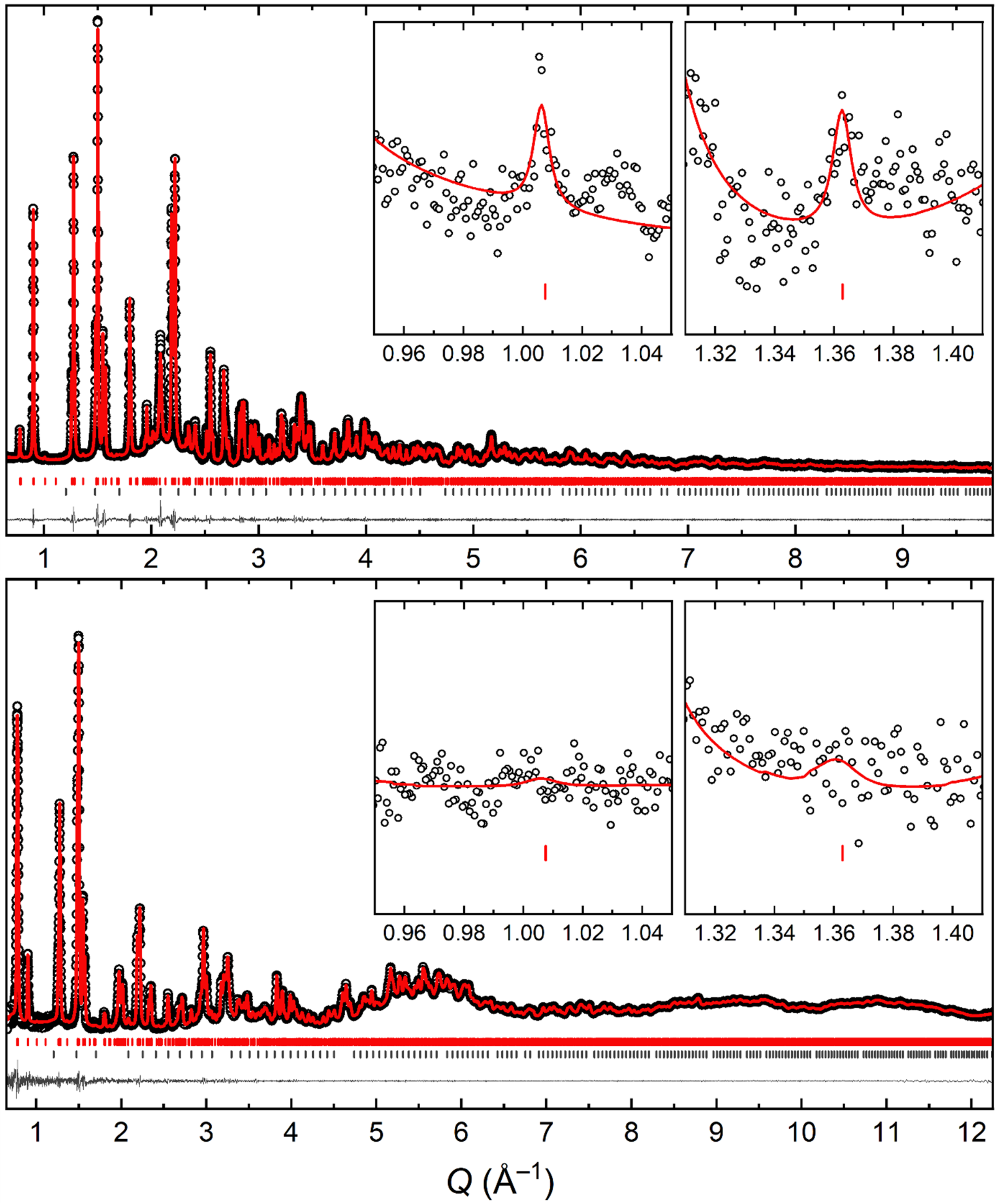

**Figure S2 │** Synchrotron X-ray (top) and neutron (bottom) powder diffraction patterns for the $Yb_2CsC_{60}$ sample obtained at 300 K, with corresponding Pawley fits to the measured data shown as the red trace, and the difference curve shown as the grey trace. The top set of red vertical tick marks indicates the *hkl* positions for $Yb_2CsC_{60}$, while the bottom set of black vertical tick marks indicates *hkl* positions from the minor $Yb_2O_3$ impurity phase. The enlarged views of selected areas in the insets show the presence of extremely weak reflections in the synchrotron X-ray data (top) that index as (111) and (021) on the orthorhombic unit cell – these are not discernible in the corresponding neutron data (bottom).

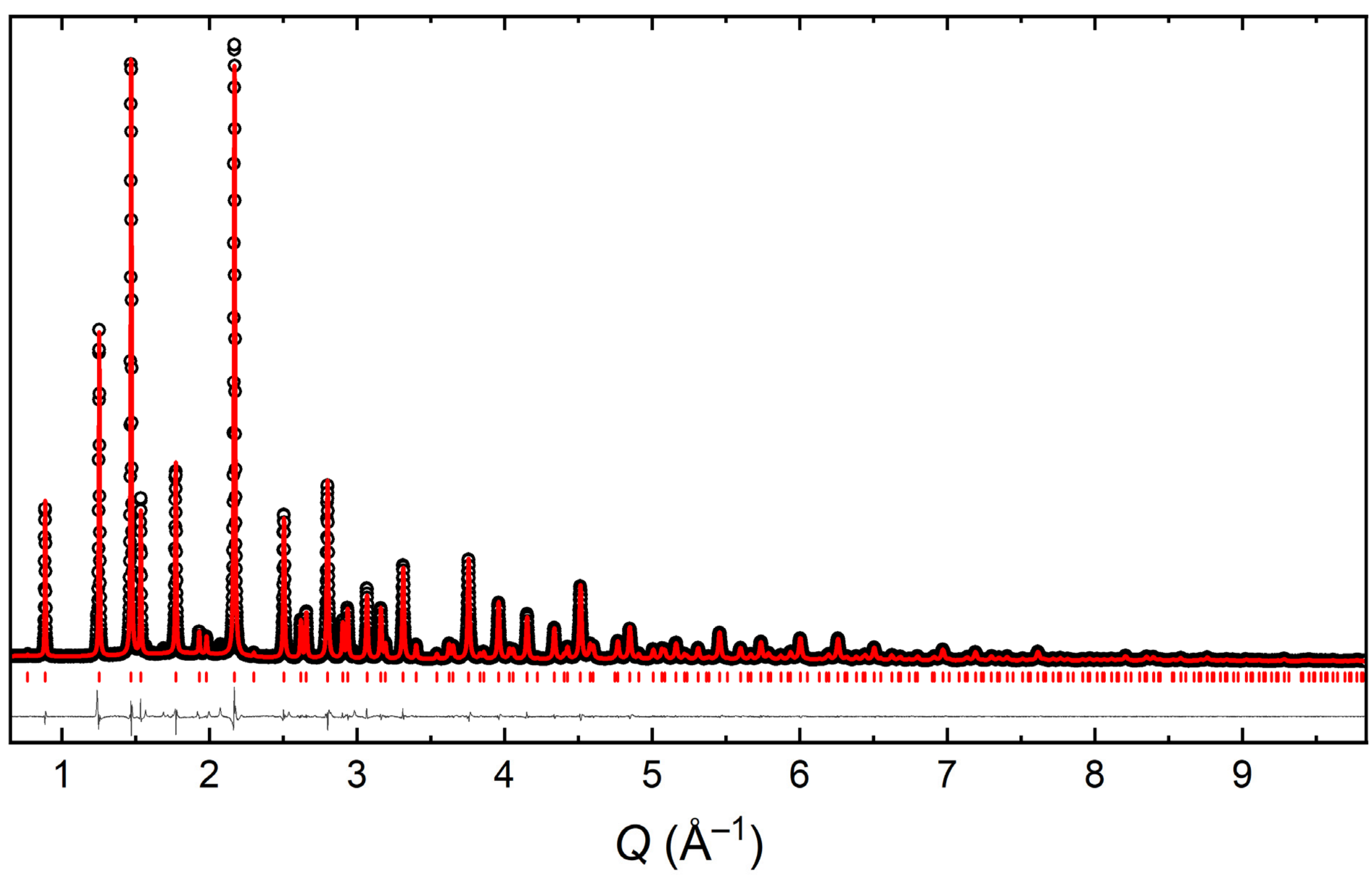

**Figure S3 │** Synchrotron X-ray powder diffraction profile for the *fcc*-structured $Ba_2CsC_{60}$ sample at ambient temperature, and the accompanying Pawley fit (BL02B2 beamline, SPring-8, Japan; λ = 0.79911 Å, space group $Fm\bar{3}$, $a$ = 14.1984(1) Å).

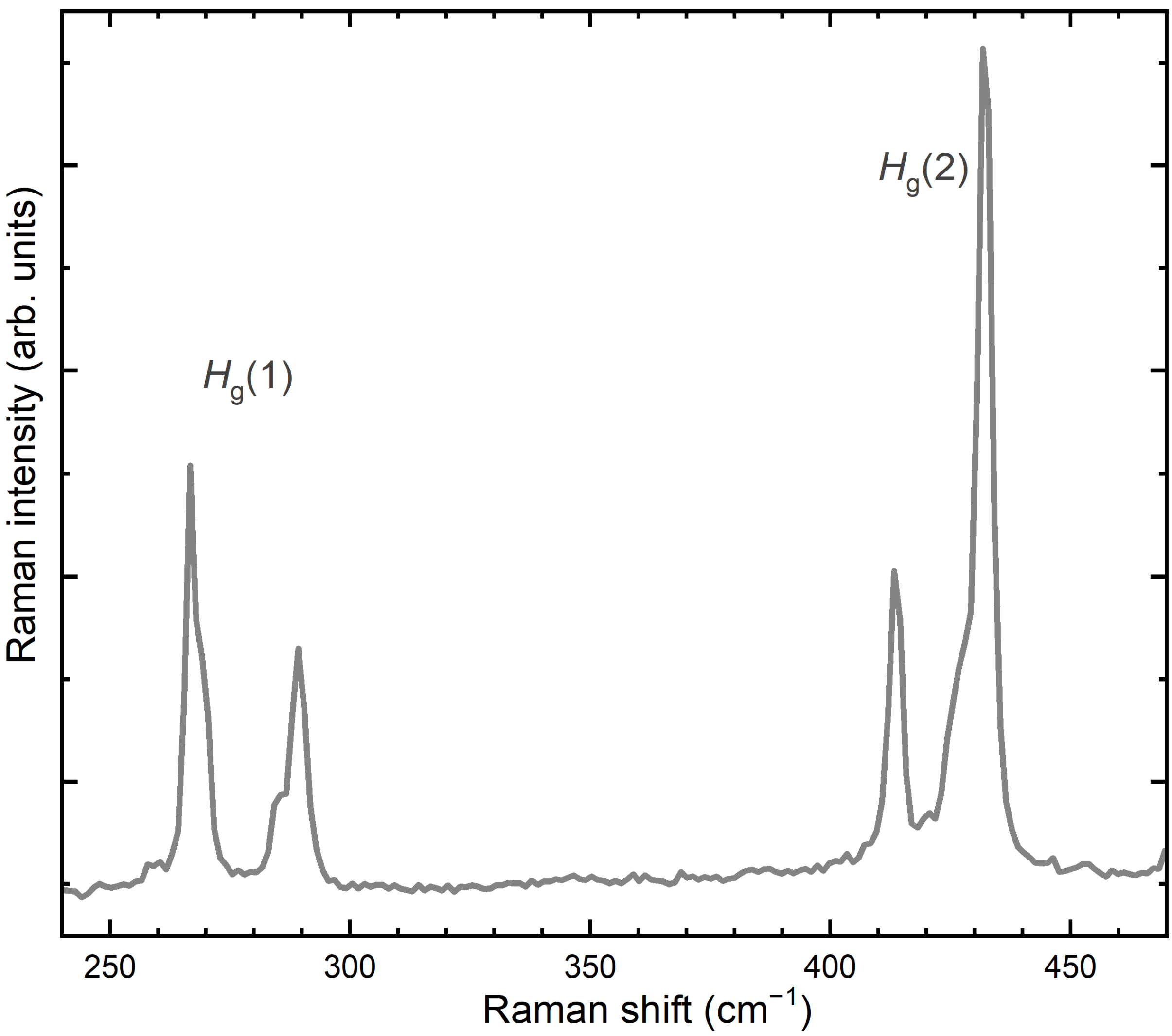

**Figure S4 │** Raman spectrum of $Yb_2CsC_{60}$ in the frequency region of the $H_g(1)$ and $H_g(2)$ modes of $C_{60}$, further supporting the lowering in symmetry observed in the powder diffraction analysis.

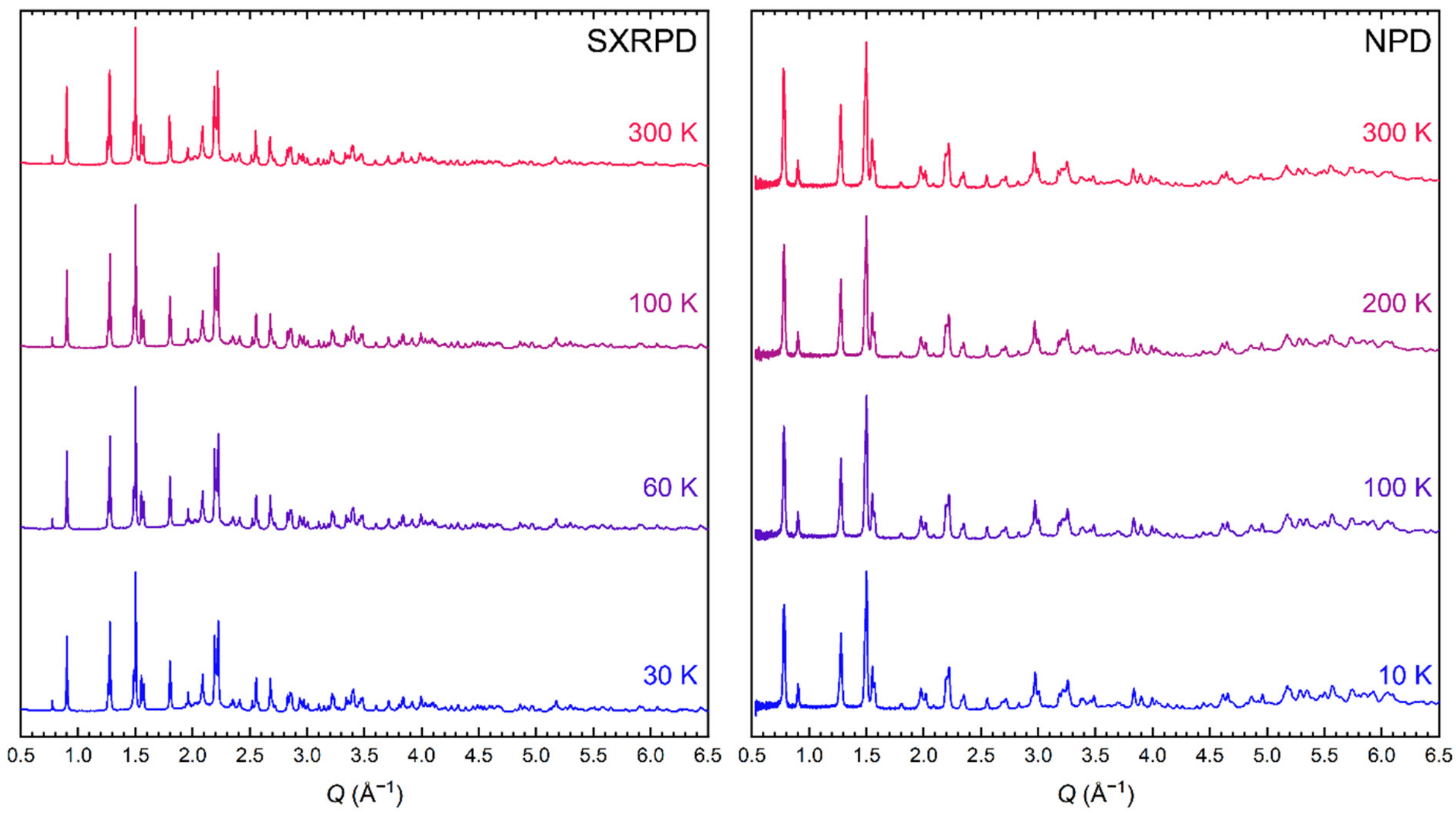

**Figure S5 │** Temperature dependence of the synchrotron X-ray powder diffraction patterns between 30 K and 300 K (left) and the neutron powder diffraction patterns between 10 K and 300 K (right) for the $Yb_2CsC_{60}$ sample.

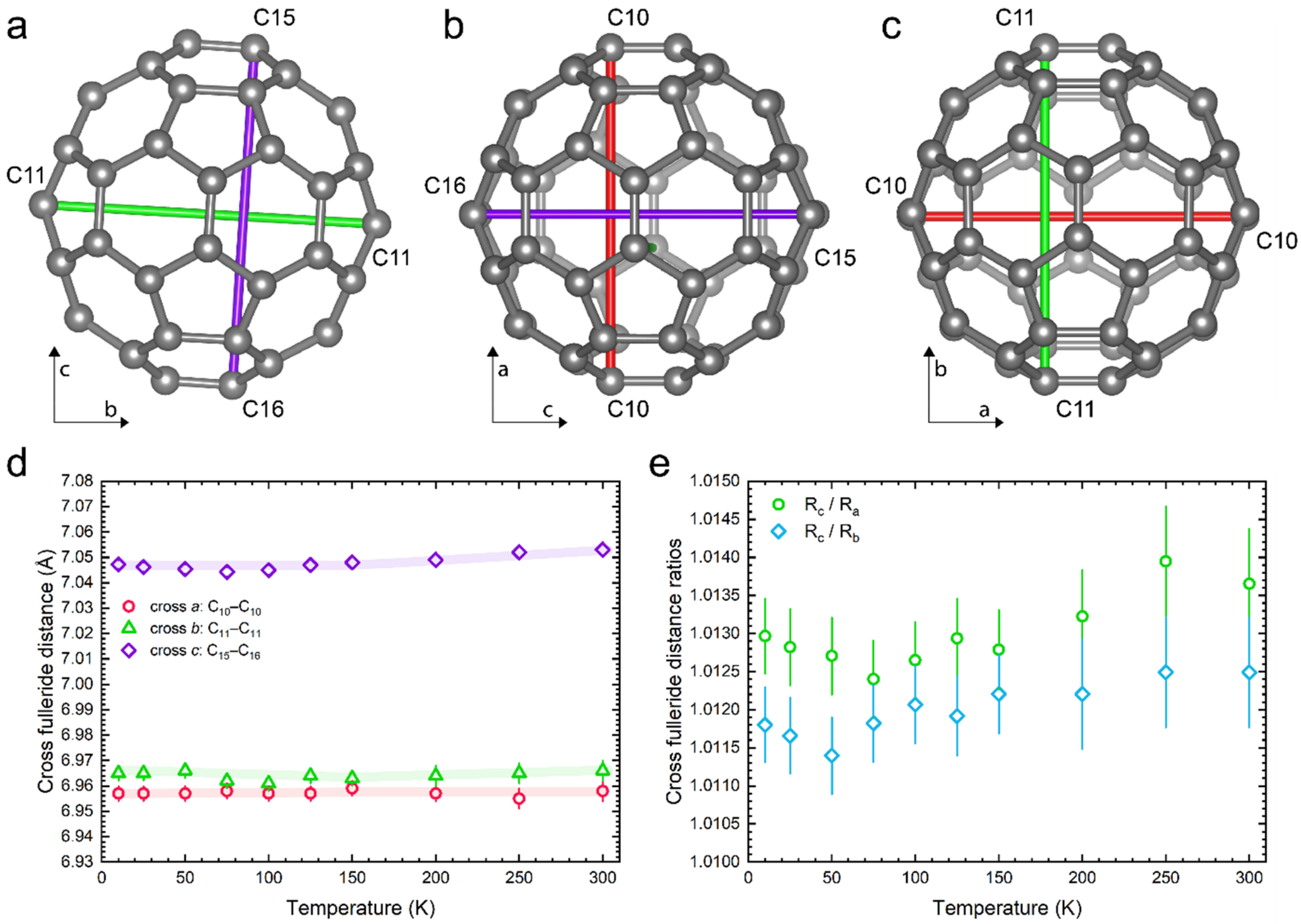

**Figure S6 │** Measurements used to determine the prolate spheroidal molecular structure of the $C_{60}^{5-}$ anion in $Yb_2CsC_{60}$. The molecular penta-anion is viewed down the crystallographic *a*- (**a**), *b*- (**b**) and *c*-axis (**c**), respectively. The cross *a*-axis distance is the diametric $C_{10}$—$C_{10}$ distance, the cross *b*-axis distance is the diametric $C_{11}$—$C_{11}$ distance, and cross *c*-axis distance is the diametric $C_{15}$—$C_{16}$ distance, as described in the main text. These distances are illustrated as the red, green, and purple contacts, respectively, and are numerically identical to the distances from mid-way along the respective bonds that lie on the symmetry axes. Cations are omitted for clarity. The distances (**d**) and resulting $^{R_c}/_{R_a}$ and $^{R_c}/_{R_b}$ ratios (**e**) are also provided in Table S4 as a function of temperature, as derived from the sequential Rietveld refinements against the NPD patterns. Unless shown, error bars are smaller than or commensurate with the symbols and represent ±1σ and are propagated from experimental uncertainties. Pale lines are guides to the eye.

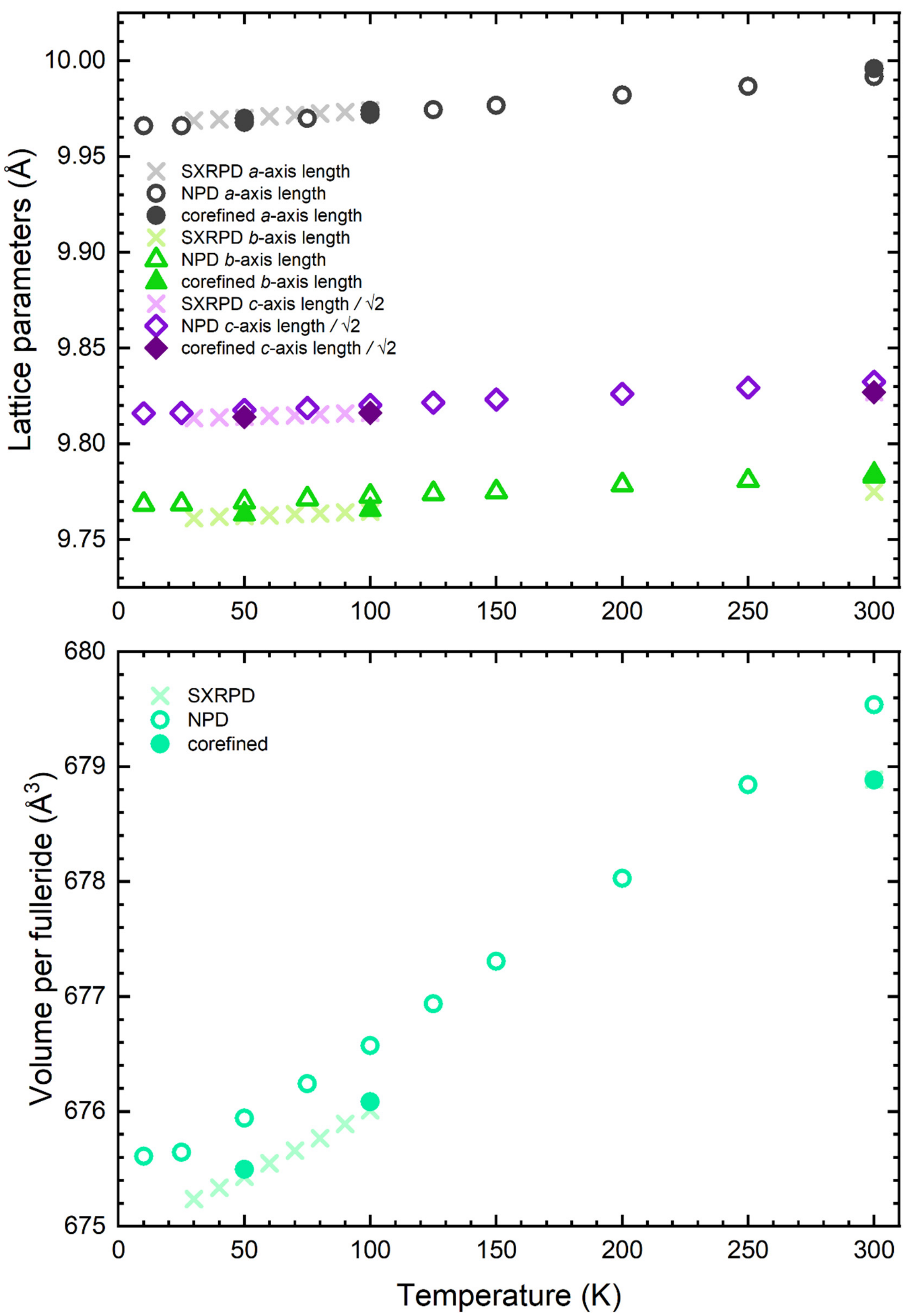

**Figure S7** │ (top) *Pmnn* orthorhombic unit cell lattice parameters, *a* (grey), *b* (green) and *c* (purple) of $Yb_2CsC_{60}$ powder as a function of temperature. The lattice parameter *c* (*ca.* 14 Å) is divided by $\sqrt{2}$ for easier comparison with the other parameters. (bottom) Unit cell volume of $Yb_2CsC_{60}$ as a function of temperature. Open symbols and cross marks represent the values extracted by Rietveld refinement of NPD and SXRPD patterns, respectively. Solid symbols show the values from co-refinements of the NPD and SXRPD data.

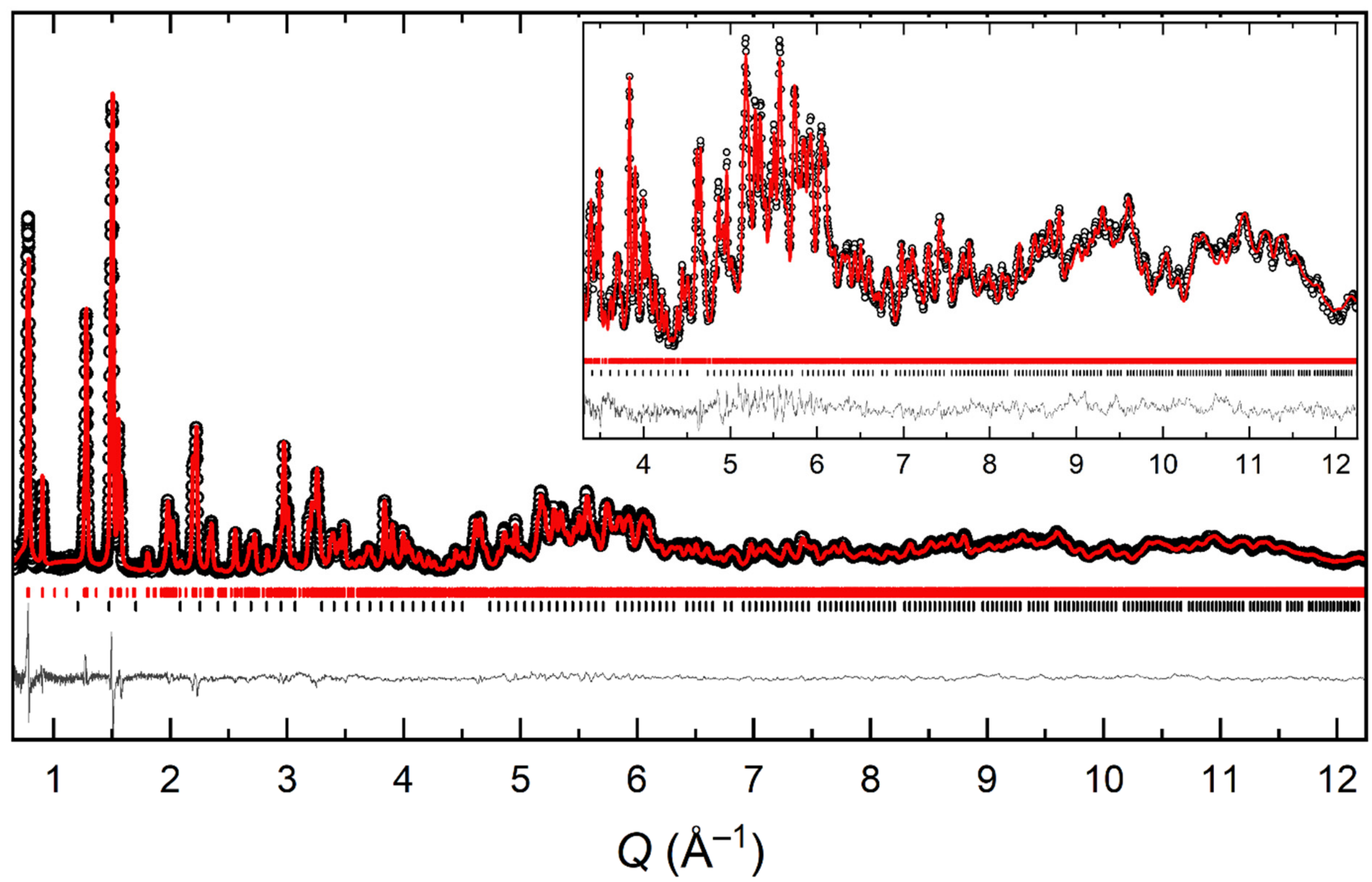

**Figure S8 │** Rietveld refinement of the neutron powder diffraction pattern obtained for $Yb_2CsC_{60}$ at 10 K. $R_{wp}$ = 4.873%, $R_p$ = 6.779%, $R_{exp}$ = 1.457%, GOF = 3.343. The black circles, red curve, grey curve, and vertical tick marks denote the raw measured data, the Rietveld refinement curve, the difference curve, and the *hkl* positions, respectively. Symbols are larger than or commensurate with their error bars which represent ±1σ.

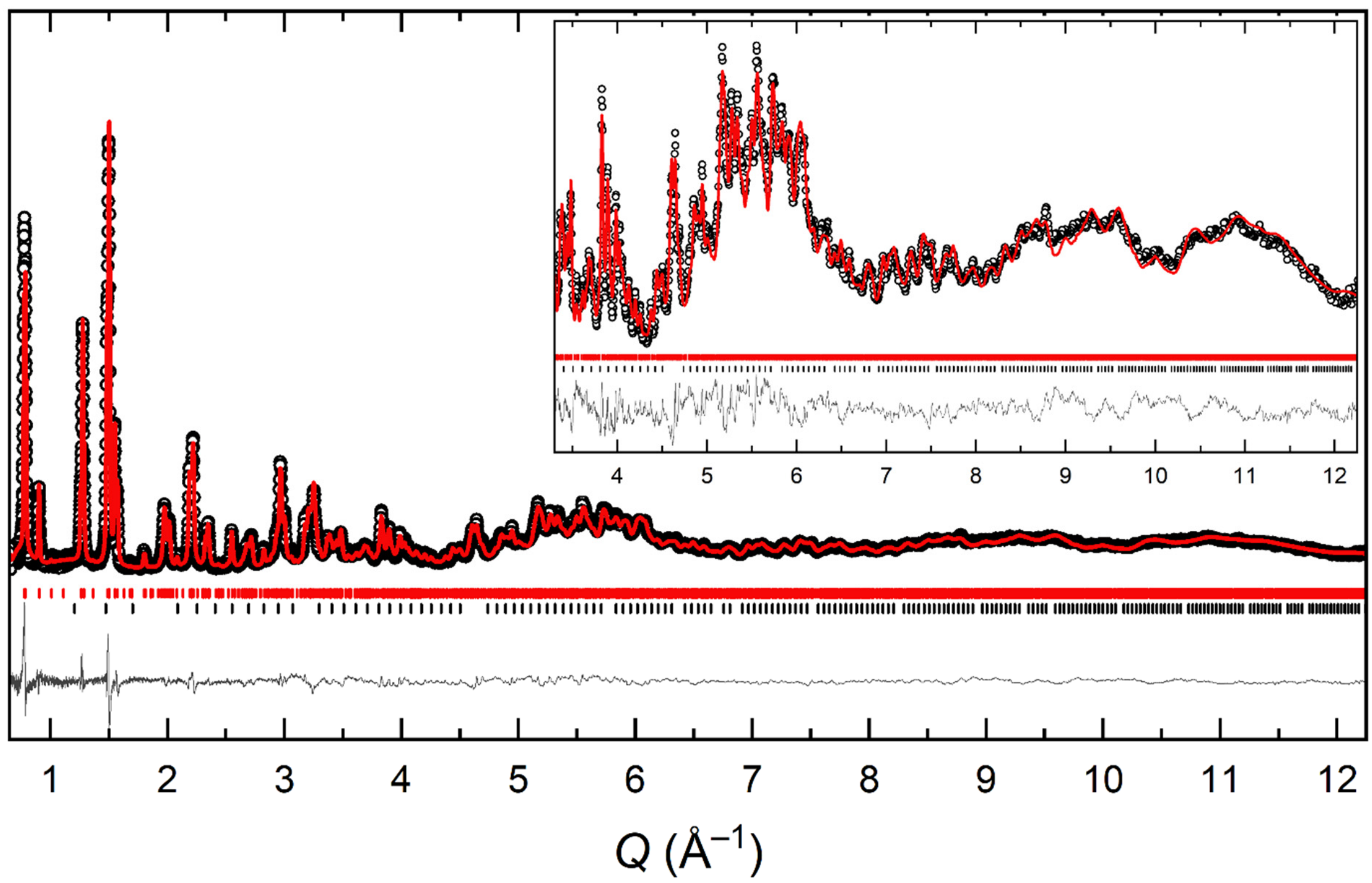

**Figure S9 │** Rietveld refinement of the neutron powder diffraction pattern obtained for $Yb_2CsC_{60}$ at 300 K. $R_{wp}$ = 6.000%, $R_p$ = 7.955%, $R_{exp}$ = 1.509%, GOF = 3.977. The black circles, red curve, grey curve, and vertical tick marks denote the raw data, the Rietveld refinement curve, the difference curve, and the *hkl* positions, respectively. Symbols are larger than or commensurate with their error bars which represent ±1σ.

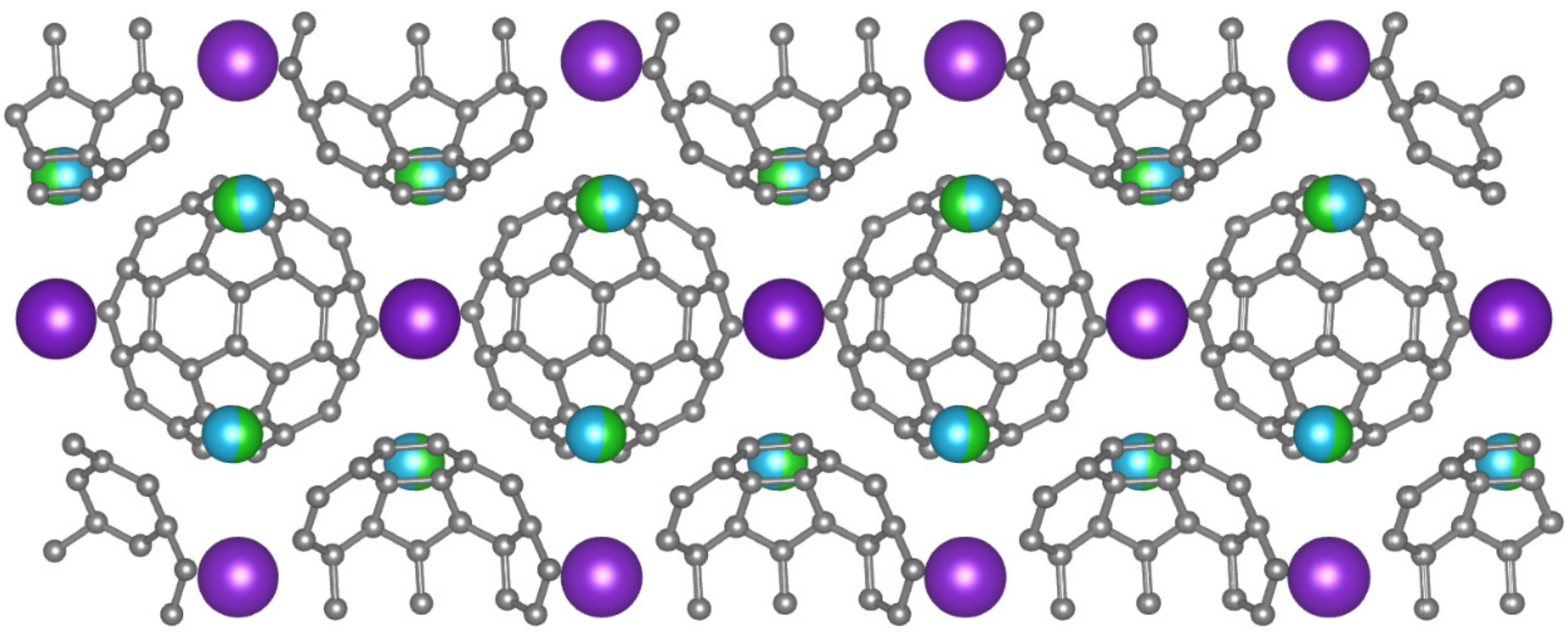

**Figure S10 │ Temperature dependence of the crystal structure of $Yb_2CsC_{60}$.** Overlayed 10 K and 300 K crystal structures for $Yb_2CsC_{60}$ from Rietveld analysis of NPD data displayed in the crystallographic *bc* plane. Purple and grey spheres depict Cs cations and C atoms, respectively. Green and teal spheres depict Yb cations at 10 and 300 K, respectively.

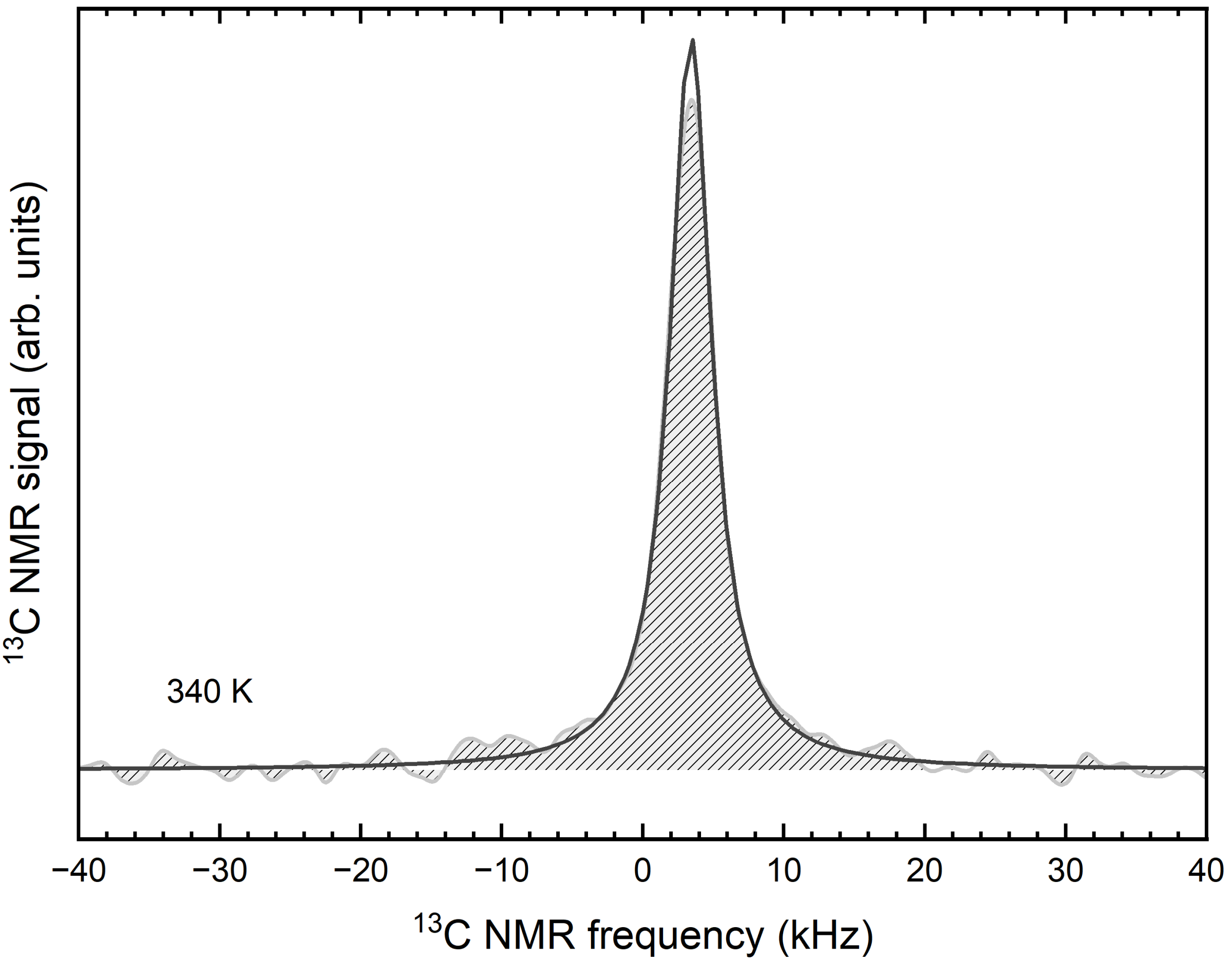

**Figure S11** │ $^{13}$C (nuclear spin $I$ = ½) NMR spectrum measured for the $Yb_2CsC_{60}$ powder sample at 340 K (light grey shaded area) at a magnetic field of 9.39 T with the corresponding Larmor resonance frequency of $^{13}\nu_{ref}$ = 100.5673 MHz. As a reference, the $^{13}$C NMR signals from adamantane were used, and the shifts were then recalculated against the tetramethylsilane (TMS) standard. The solid dark grey line is a fit to a Lorentzian lineshape.

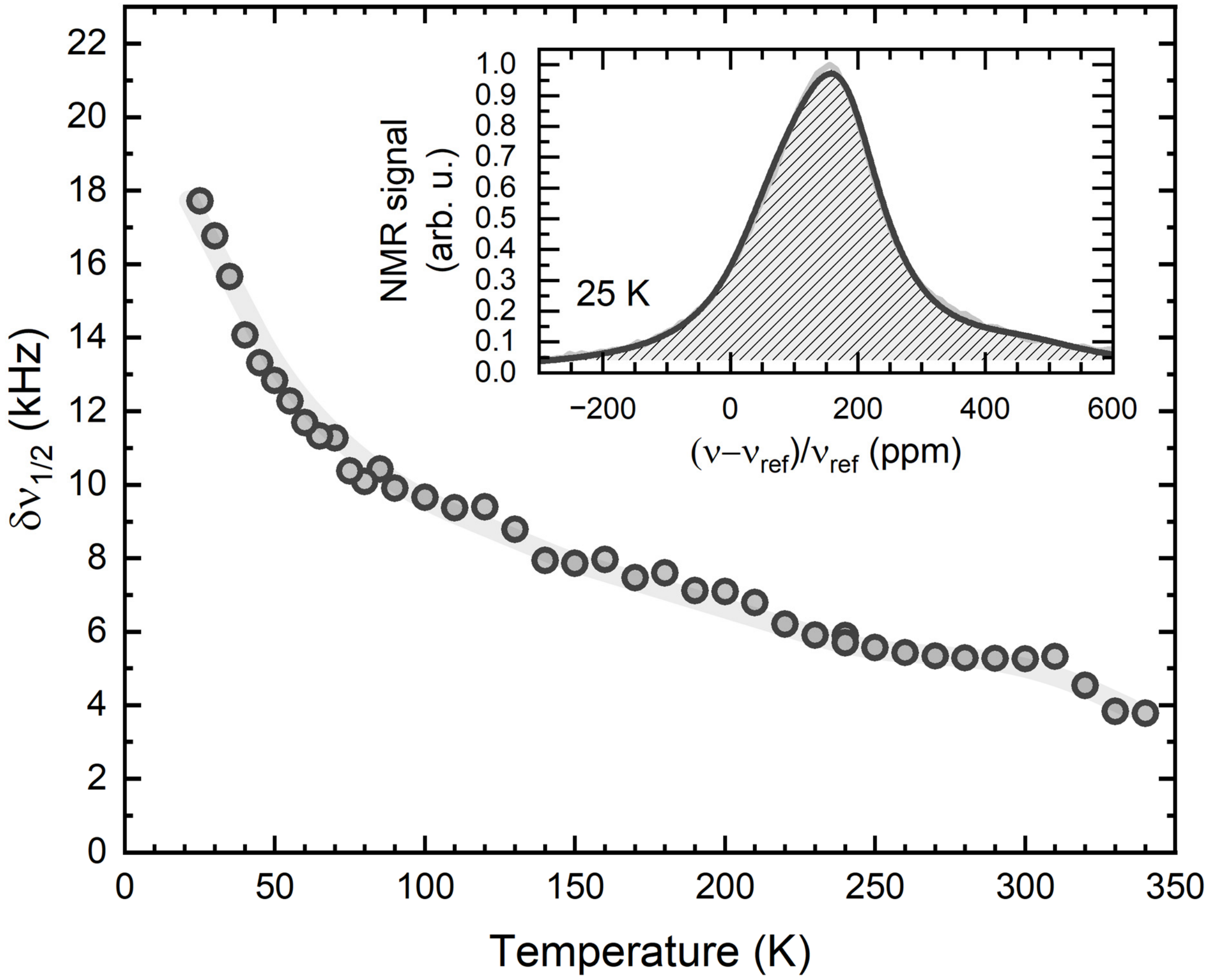

**Figure S12 │** Temperature dependence of the linewidth of $^{13}$C powder NMR spectra of $Yb_2CsC_{60}$ (grey circles). *Inset*: $^{13}$C NMR spectrum measured at 25 K in a magnetic field of 9.39 T with the corresponding Larmor resonance frequency, $^{13}\nu_{ref}$ = 100.5673 MHz (light grey shaded area) together with a lineshape fit (solid dark grey line) comprising of a dominant axially symmetric powder lineshape and minor shifted component. The solid line is a guide to the eye.